\documentclass[prl,twocolumn,showpacs,floatfix,superscriptaddress]{revtex4-1}%
\usepackage{amsmath,amssymb,amsfonts}
\usepackage{longtable}
\usepackage{graphicx}
\usepackage{bbm}
\usepackage{enumitem}
\usepackage{esint}
\usepackage[colorlinks=true,     linkcolor=blue,     citecolor=blue,
urlcolor=black]{hyperref}
\usepackage{MnSymbol}
\usepackage{amsmath}
\usepackage{amsfonts}
\usepackage{amssymb}
\usepackage{umoline}
\usepackage{color}

\usepackage{graphicx}
\usepackage{epsfig}
\usepackage{amsmath,bbm}
\usepackage{amsfonts,amssymb}
\usepackage{dsfont}

\usepackage{cancel}

\newcommand{\vspacefortable}{\vspace*{0.3cm}}

\newcommand{\sabs}[1]{|#1|}
\newcommand{\ket}[1]{\left|#1\right\rangle}
\newcommand{\bra}[1]{\left\langle#1\right|}
\newcommand{\braket}[1]{\left\langle#1\right\rangle}

\newcommand{\mO}{\mathcal{O}}
\newcommand{\mA}{\mathcal{A}}
\newcommand{\mI}{\mathcal{I}}

\newcommand{\mH}{\mathcal{H}}

\newcommand{\mL}{\mathcal{L}}

\newcommand{\mP}{\mathcal{P}}
\newcommand{\mQ}{\mathcal{Q}}
\newcommand{\mD}{\mathcal{D}}

\newcommand{\rmd}{\textrm{d}}

\newcommand{\hc}{\textrm{h.c.}}

\newcommand{\mean}[1]{\left\langle #1 \right\rangle}

\newcommand{\hHint}{{H}_{\mathrm{int}}}
\newcommand{\Oopd}{\hat{\mO}^\dagger}
\newcommand{\Oop}{\hat{\mO}}
\newcommand{\hV}{\hat{V}}

\newcommand{\vect}[1]{\mathbf{#1}}

\newcommand{\bal}{\begin{equation}}
\newcommand{\eal}{\end{equation}}

\newcommand{\lat}{{\llangle}}
\newcommand{\rat}{{\rrangle}}

\newcommand{\eeqref}[1]{Eq.\,\eqref{#1}}
\newcommand{\figref}[1]{Fig.\,\ref{#1}}

\begin{document}

\title{Constrained dynamics via the Zeno effect in quantum simulation:\\Implementing non-Abelian lattice gauge theories with cold atoms}
\author{K. Stannigel}
\email{Equal contribution.}
\affiliation{Institute for Quantum Optics and Quantum Information of the Austrian
Academy of Sciences, 6020 Innsbruck, Austria}
\author{P. Hauke}
\email{Equal contribution}
\email{philipp.hauke@uibk.ac.at}
\affiliation{Institute for Quantum Optics and Quantum Information of the Austrian
Academy of Sciences, 6020 Innsbruck, Austria}
\author{D. Marcos}
\affiliation{Institute for Quantum Optics and Quantum Information of the Austrian
Academy of Sciences, 6020 Innsbruck, Austria}
\author{M. Hafezi}
\affiliation{Joint Quantum Institute, NIST / University of Maryland, 20742 College Park MD, USA}
\author{S. Diehl}
\affiliation{Institute for Quantum Optics and Quantum Information of the Austrian
Academy of Sciences, 6020 Innsbruck, Austria}
\affiliation{Institute for Theoretical Physics, University of Innsbruck, 6020 Innsbruck, Austria}
\author{M. Dalmonte}
\email{marcello.dalmonte@uibk.ac.at}
\affiliation{Institute for Quantum Optics and Quantum Information of the Austrian
Academy of Sciences, 6020 Innsbruck, Austria}
\affiliation{Institute for Theoretical Physics, University of Innsbruck, 6020 Innsbruck, Austria}
\author{P. Zoller}
\affiliation{Institute for Quantum Optics and Quantum Information of the Austrian
Academy of Sciences, 6020 Innsbruck, Austria}
\affiliation{Institute for Theoretical Physics, University of Innsbruck, 6020 Innsbruck, Austria}

\begin{abstract}
We show how engineered classical noise can be used to generate
constrained Hamiltonian dynamics in atomic quantum simulators of many-body
systems, taking advantage of the continuous Zeno effect. After discussing the general theoretical framework, we focus on
applications in the context of lattice gauge theories, where imposing exotic,
quasi-local constraints is usually challenging. We demonstrate the
effectiveness of the scheme for both Abelian and non-Abelian gauge theories, and discuss how engineering dissipative constraints substitutes complicated, non-local interaction patterns by global coupling to laser fields.
\end{abstract}

\pacs{03.65.Xp,37.10.Jk,11.15.Ha}

\maketitle

Laboratory experiments with atomic quantum-degenerate gases have established a
synergetic link between atomic physics and condensed matter \cite{Lewenstein2012,Cirac2012,Bloch2012}, and hold prospects for a similar connection to high-energy physics \cite{Kapit2011,Banerjee2012,Tagliacozzo2012,Zohar2012,Kasamatsu2012,Banerjee2013,Tagliacozzo2013,Zohar2013c,Zohar2013a}. 
Loaded into optical lattices, cold atoms realize Hubbard models,
which can be designed and controlled via external fields to mimic the dynamics
of quantum many-body systems in equilibrium and non-equilibrium situations~\cite{Bloch2008,Bloch2012}. 
While a focus of research during the last decade has been the development of a
toolbox for designing specific lattice Hamiltonians \cite{Lewenstein2012,Cirac2012,Bloch2012}, we address below the problem of implementing desired Hubbard dynamics \emph{in the presence of constraints}, i.e., we wish to keep the
system dynamics within a certain subspace of the total Hilbert space. A
familiar way of imposing such constraints is to add an energy penalty to the
Hamiltonian~\cite{Lacroix2010}. Below, we describe an alternative scenario that is based on driving the
system with engineered classical noise, exploiting the Zeno effect \cite{Viola1999,Wu2002b,Beige2000,Facchi2008,Wu2009,Raimond2010}. 
As we will see, `adding noise'
provides a general tool to implement -- in an experimentally efficient and
accessible way -- highly nontrivial constraints in quantum many-body systems. 

\begin{figure}[ptb]
\includegraphics[width=0.42\textwidth]{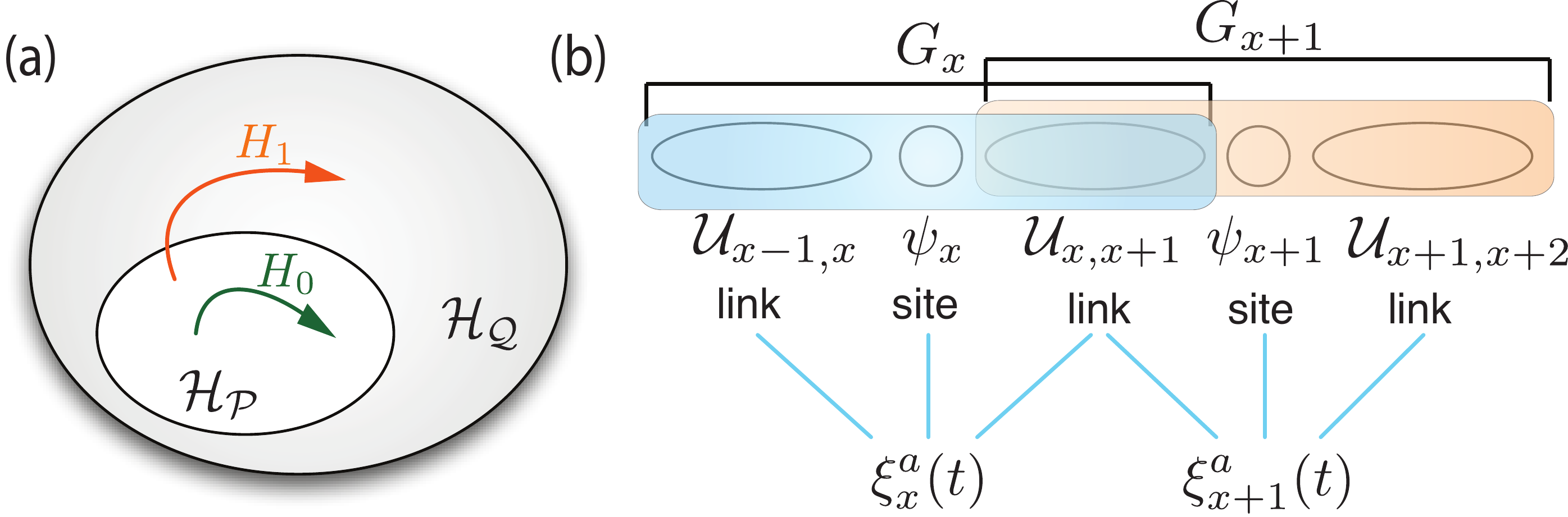}
\caption{
Dissipative protection of gauge invariance in implementations of a lattice gauge theory (LGT).  
(a) The dynamics $H_0$ happens within the physically relevant subspace $\mathcal H_{\mathcal{P}}$, defined by $G_{x}^{a}|\psi\rangle=0$, 
but gauge-variant perturbations $H_1$ may drive the system into the unphysical subspace $\mathcal H_{\mathcal{Q}}$ (where $G_{x}^{a}|\psi\rangle\neq 0$). 
(b) LGTs consist of fermions $\psi_x$ living on sites, coupled to gauge fields $\mathcal{U}_{x,x+1}$ living on links. 
The dynamics can be constrained to the physical subspace by coupling independent noise sources linearly to each generator. 
The multi-site structure of the generators implies that the noise has to be correlated quasi-locally in space. 
}%
\label{fig:1}%
\end{figure}

The present work is motivated by the ongoing quest to build a quantum
simulator for Abelian and non-Abelian lattice gauge theories (LGTs) with cold atoms in optical
lattices \cite{Banerjee2012,Tagliacozzo2012,Zohar2012,Kasamatsu2012,Banerjee2013,Tagliacozzo2013,Zohar2013c,Zohar2013a}. LGTs play a prominent role in both particle and condensed matter
physics: in the standard model, the interaction between constituents of matter
are mediated by gauge bosons \cite{Montvay1994,Creutz1997,DeGrand2006,Gattringer2010}, and in frustrated magnetism, quantum spin
liquids are suitably described in the language of gauge theories \cite{Kogut1979,Lee2006,Lacroix2010}. The key
feature of a LGT is the presence of local (gauge) symmetries. The generators
$G_{x}^{a}$ of these local gauge transformations, with $x$ denoting lattice
sites and $a$ a color index, commute with the lattice
Hamiltonian, $\left[  H_{0},G_{x}^{a}\right]  =0$ for all $x,a$, and thus
provide local conservation laws. 
They can be interpreted in analogy to Gauss's law from electrodynamics, as
they constrain the dynamics of the system to a \emph{physical subspace}
$\mathcal{H}_{\mathcal{P}}$ given by the constraints $G_{x}^{a}|\psi\rangle=0$~\cite{Fradkin_book}. 
In
high-energy physics, gauge symmetries are from fundamental considerations exact, 
but in quantum simulations these symmetries will normally be approximate on some level in the microscopic model. Thus, the microscopic Hamiltonian will be of the form
$H_{\mathrm{micro}}=H_{0}+H_{1}$, with $H_{0}\sim J$ the desired gauge-invariant part, and
$H_{1}\sim\lambda$ a perturbation, which drives the system dynamics outside of the gauge-invariant subspace $\mathcal{H}_{\mathcal{P}}$, see Fig.~\ref{fig:1}(a). 
A central challenge of quantum simulating LGTs is to introduce mechanisms that suppress these errors.

\emph{Gauge constraints via classical noise.} 
A common strategy to restrict the dynamics to a certain subspace is by adding an \textit{energy penalty} to the microscopic Hamiltonian: $H_{\rm micro}= H_0+H_1+H_U$, with $H_U \sim U \gg \lambda$. In this case, to order $\lambda^2/U$, manifolds with different eigenvalues of $H_U$ become decoupled. In the context of LGTs, one can use $H_U = U \sum_{x,a} (G_x^a)^2$ to impose the Gauss law constraints \cite{Lacroix2010,Banerjee2012,Zohar2012,Banerjee2013,Marcos2013,Hauke2013b}. 
Since in a LGT the generators $G_{x}^{a}$ are complicated expressions of the matter and gauge fields -- especially in non-Abelian models -- adding an interaction term involving the \emph{square} of these poses a formidable challenge \cite{Note1}.
In contrast, we pursue here the strategy of enforcing the constraints $G_{x}%
^{a}|\psi\rangle=0$ by adding \emph{classical noise terms} \cite{footnoteLangevinComputation,Batrouni1985} to the microscopic Hamiltonian
\begin{equation}
\label{eq:Hclassicalnoise}
H_{\rm micro}(t)=H_{0}+H_{1}+\sqrt{2\kappa}\sum_{x,a}\xi_{x}^{a}(t)G_{x}^{a}\,, 
\end{equation}
which are \emph{linear} in $G_{x}^{a}$ and involve independent white-noise
processes $\xi_{x}^{a}(t)$ with
$\llangle{\xi_x^a(t) \xi_y^b(t^\prime)}\rrangle=\delta_{xy}\delta_{ab}%
\delta(t-t^{\prime})$. 
Each realization of the noise will give rise to an evolution of the system described by a stochastic state vector $|{\psi
(t)}\rangle$. Averaging over the noise fluctuations leads to the density
operator $\rho=\llangle{\ket{\psi(t)}\bra{\psi(t)}}\rrangle$, obeying the master equation (see supplemental material \cite{supp_dissipative-protection})
\begin{align}
\label{eq:ME}
\dot{\rho}=-iH_{\mathrm{eff}}\rho+i\rho H_{\mathrm{eff}}^{\dag}+2\kappa
\sum_{x,a}G_{x}^{a}\rho G_{x}^{a}\,,
\end{align}
with non-Hermitian Hamiltonian $H_{\mathrm{eff}}$, 
\[
H_{\mathrm{eff}}=H_{0}+H_{1}-i\kappa\sum_{x,a}(G_{x}^{a})^{2}\,.
\]
The effective Hamiltonian $H_{\mathrm{eff}}$ contains a damping term involving the
\emph{square} of the generators, introduced by the noisy \emph{single-particle} terms in Hamiltonian \eqref{eq:Hclassicalnoise}. 
For $\kappa/\lambda\gg 1$, this term constrains the
evolution to $\mathcal{H}_{\mathcal{P}}$. In fact, this scale separation of a large rate vs.\ a small energy scale provides the dissipative analogue of the energetic protection described above, which relies on the separation of two energy scales. 
In the present case, the protection term arises from a linear coupling of the generators to a classical noise source, which -- in contrast to an energy penalty $H_U$ -- does not require complicated, non-local two-body interactions.

To demonstrate the dissipative protection explicitly, we integrate out the
dynamics of the gauge-variant space $\mathcal{H}_{\mathcal{Q}}$, defined by  $G_{x}^{a}|\psi\rangle\neq 0$, 
to leading order in $\lambda/\kappa$. Denoting by $\mathcal{P}$ the projector on $\mathcal{H}%
_{\mathcal{P}}$, we obtain the following master equation for the
gauge-invariant part of the density operator $\rho_{\mathcal{P}\mathcal{P}%
}=\mathcal{P}\rho\mathcal{P}$
\[
\dot{\rho}_{\mathcal{P}\mathcal{P}}=-i\tilde{H}_{\mathrm{eff}}\,\rho_{\mathcal{P}\mathcal{P}}+i\rho_{\mathcal{P}\mathcal{P}}\,\tilde
{H}_{\mathrm{eff}}^{\dag}\,, \label{eq:master2}
\]
with effective non-Hermitian Hamiltonian
\begin{equation}
\tilde{H}_{\mathrm{eff}}\approx\mathcal{P}(H_{0}+H_{1})\mathcal{P}-i\,\mathcal{P}%
H_{1}\mathcal{Q}\frac{1}{\kappa\sum_{x,a}(G_{x}^{a})^{2}}\mathcal{Q}%
H_{1}\mathcal{P}\,, \label{eq:Heff2}%
\end{equation}
where $\mathcal{Q}=\mathbbm{1}-\mathcal{P}$ \cite{supp_dissipative-protection}. Starting from an initial state $|{\psi_{0}}\rangle\in\mathcal{H}_{\mathcal{P}}$, the evolution under this Hamiltonian is restricted to $\mathcal{H}_{\mathcal{P}}$, but reduces the norm of the state vector, signifying the transfer of population to the gauge-variant subspace $\mathcal{H}_{\mathcal{Q}}$. 
This population transfer sets a timescale $t\lesssim \kappa/\lambda^2$ below which the dissipative protection of local quantities is effective. 
Therefore, in the strong-noise limit considered in this work,  gauge invariance is protected for times long compared to the time scales at which errors accumulate without the engineered noise, $t\sim 1/\lambda$.

This suppression of transitions by fast classical fluctuations is related to motional narrowing \cite{Milburn1988,Cohen-Tannoudji1992}, and to dynamical decoupling techniques \cite{Viola1998} such as bang-bang control \cite{Viola2000,Wu2002b}, which  suppress unwanted couplings to an environment, e.g., in a quantum information context. More specifically, our scheme can be seen as a classical analogue of the quantum Zeno effect  \cite{Kofman2001a, Facchi2008}, which has been discussed in the context of quantum control \cite{Wu2009,Beige2000,Raimond2010}, but also of quantum many-body systems 
\cite{Syassen-2008, Daley2009,Kantian2009,Roncaglia2010}.
In the standard quantum Zeno effect, the required dissipation originates from an interaction of the system with quantum fluctuations of the bath or frequent measurements, while here the dissipation is simulated by classical fluctuations of the perturbation field.

\begin{figure}[t]
\includegraphics[width=0.42\textwidth]{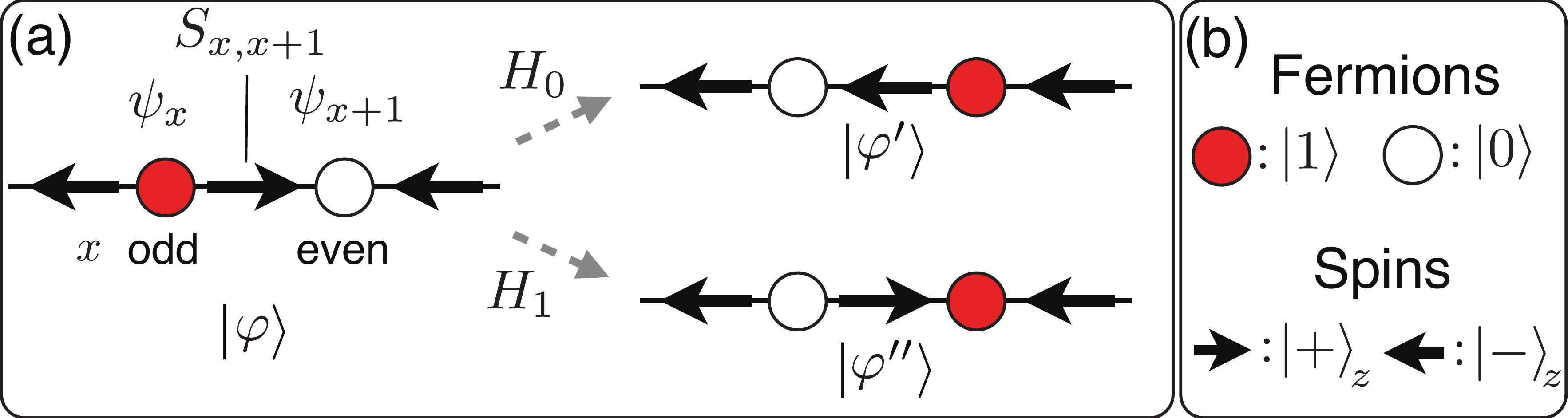}\\
\vspace*{0.2cm}
\includegraphics[width=0.44\textwidth]{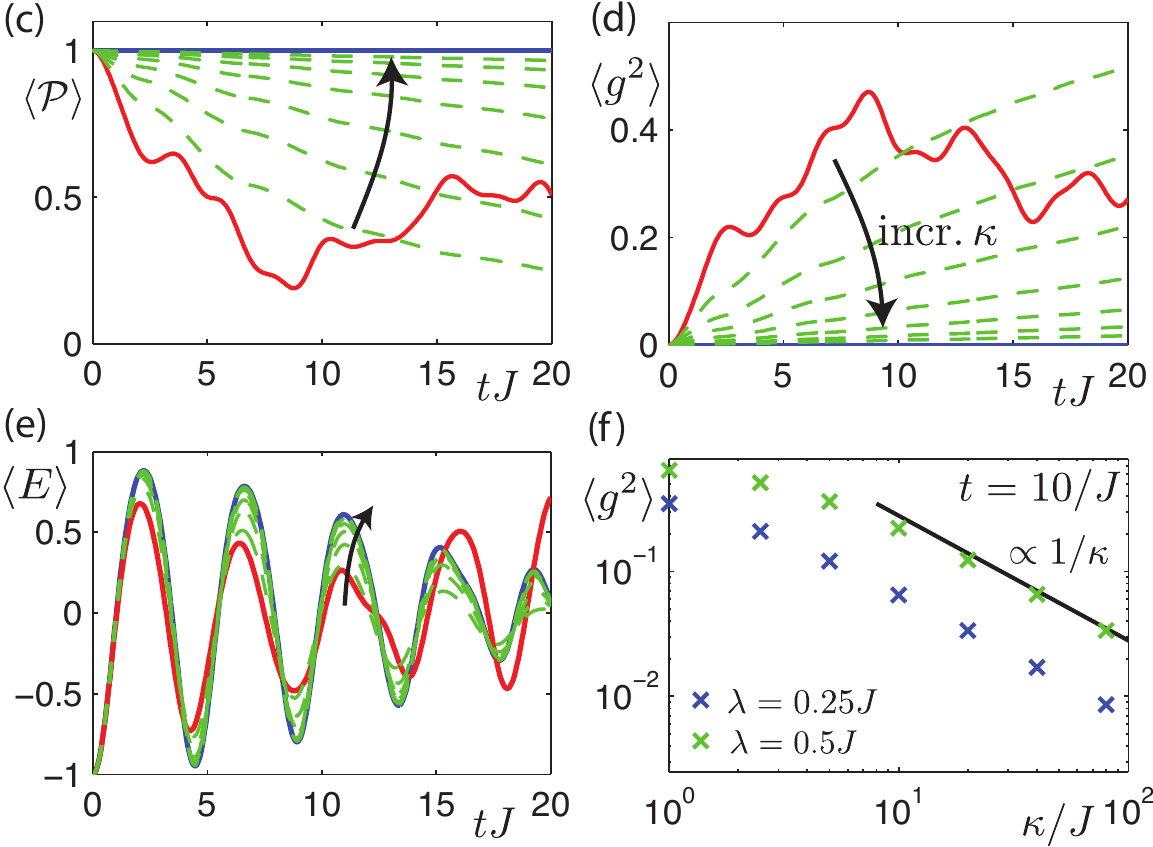}\caption{
(a) System dynamics of a U(1) LGT in the ideal case ($H_0$, top) and under the effect of undesired single particle tunneling ($H_1$, bottom); (b) on-site Hilbert spaces. 
(c-f) Dissipative
protection of quench dynamics in a $U(1)$ QLM with $N_{s}=4$
sites connected by $N_{l}=3$ links (open boundary conditions). 
(c) Population of gauge-invariant subspace. 
(d,f) Average violation of gauge constraint, quantified by $g^{2}=\sum_{x} G^{2}_{x}/N_{s}$.
(e) Average electric field
$E=\sum_{x} E_{x,x+1}/N_{l}$. 
In panels (c-e), blue curves indicate the
ideal dynamics. 
Red curves show the detrimental influence of gauge-variant
fermion tunneling ($\lambda/J=0.25$). 
The arrows show how increasing $\kappa$ restores the ideal dynamics (green curves; $\kappa/J=1,2.5,5,10,20,40,80$). 
Panel (f) shows the scaling with $\kappa$ of $\langle g^{2} \rangle$ at a fixed
time; the black line is a guide to the eye indicating a scaling $\propto
1/\kappa$. 
All results are obtained from the full master equation with Hamiltonian $H_0+H_1$ (see text), starting from the eigenstate of $H_0$ for $m\rightarrow\infty$.
}%
\label{fig:2quench}%
\end{figure}

\emph{Dissipative protection in Abelian LGTs.} 
As a first illustrative
and conceptually simple example, we demonstrate protection of gauge invariance in a one-dimensional U(1) lattice model, the Schwinger model, whose Hamiltonian takes the form \cite{Wiese2013}
\begin{equation}
H_{0}=\sum_{x}[J(\psi_{x}^{\dagger}\mathcal{U}_{x,x+1}\psi_{x+1}%
+\text{h.c.})+m(-1)^{x}\psi_{x}^{\dagger}\psi_{x}+\frac{\tilde{g}^{2}}{2}E_{x,x+1}%
^{2}]. \label{H_U1}%
\end{equation}
Here, $\psi_{x}$ are (staggered) fermionic \textit{matter} fields defined on the
vertices of the lattice, $\mathcal{U}_{x,x+1}$ are the \textit{gauge} fields
defined on the bonds between $x$ and $x+1$ (see Fig.\ \ref{fig:1}(b)), and $E_{x,x+1}$ is the
corresponding electric field, satisfying $[E_{x,x+1},\mathcal{U}%
_{x,x+1}]=\mathcal{U}_{x,x+1}$ \cite{Montvay1994,Creutz1997,DeGrand2006,Gattringer2010}. 
The potential term $\sim m$ corresponds to a mass term for the fermionic fields, 
whose alternating sign stems from the use of `staggered fermions'~\cite{Kogut1979}; 
$J$ and $\tilde{g}$ are the tunneling and gauge-coupling coefficients, respectively. The generators of the $U(1)$ gauge
transformations for this model are given by $G_{x}=\psi_{x}^{\dag}\psi
_{x}-E_{x,x+1}+E_{x-1,x}+[(-1)^{x}-1]/2$.
The corresponding Gauss law $G_{x}|\psi\rangle=0$ is the lattice equivalent of the one of continuum quantum electrodynamics. 
In the Wilson formulation of LGTs \cite{Wilson1974, Kogut1979}, $\mathcal{U}_{x,x+1}$ are complex
phase variables, but for our purposes the {\it quantum link model} (QLM) formalism \cite{Horn1981,Orland1990,Chandrasekharan1997,Brower1999,Wiese2013} is more convenient, where the link variables are represented by spin degrees of freedom, i.e., $\mathcal{U}_{x,x+1}\equiv S_{x,x+1}^{+},E_{x,x+1}%
\equiv S_{x,x+1}^{z}$. 
We choose here a representation using spin-$1/2$ degrees of freedom, which corresponds to a 1D version of the Schwinger model with a finite electric flux running through the system, also known as finite $\theta$-angle~\cite{Banerjee2012,Wiese2013}. 
Despite its semplicity, this model displays various interesting features related to gauge theories, such as confinement and string-breaking phenomena~\cite{Banerjee2012,Wiese2013}. 

The system dynamics between two sites as induced by Eq.~\eqref{H_U1} is sketched in Fig.~\ref{fig:2quench}(a-b). On the left-hand side, a typical gauge-invariant state $|\varphi\rangle$ is illustrated, where the Gauss law is satisfied at both vertices. Under the action of the correlated tunneling contained in $H_0$, $(\psi^\dagger_{x+1}S^-_{x,x+1}\psi_{x}+\textrm{h.c.})$, the fermion tunnels from $x$ to $x+1$, and the center spin $\vec{S}_{x, x+1}$ flips, preserving gauge invariance in the final state $|\varphi'\rangle$. Processes of this kind describe the creation of a particle-antiparticle pair accompanied by an excitation of the gauge field.
In typical implementations, additional gauge-variant imperfections appear, such as single-fermion tunneling $H_{1}=\lambda\sum_{x}(\psi_{x+1}^{\dag}\psi_{x}+\mathrm{h.c.})$. 
Once such processes are allowed, the system dynamics involves states of the form $|\varphi''\rangle$, where the condition $G_{x}|\varphi''\rangle=0$ is not satisfied for all $x$, thus leading to leakage into $\mathcal{H}_{\mathcal{Q}}$. Imperfections of this kind correspond to a creation of a particle-antiparticle pair without any effect on the gauge fields, destroying gauge invariance.

To illustrate the dissipative protection in this model, we study quantum-quench dynamics as illustrated in Fig.~\ref{fig:2quench}(c-f), where we prepare the system in the ground state of $m=\infty$ and quench to $m=0$ at time $t=0$.
The system evolves under $H_0$ given by Eq.~\eqref{H_U1} plus a gauge-variant fermion hopping $H_{1}$ \cite{footnoteFigureOfMerit}.
For $\lambda\neq0$, gauge invariance is clearly violated and the evolution of observables deviates from the ideal case (red curves in panels (c-e)). 
Increasing the strength $\kappa$ of the dissipation (green curves) gradually restores the ideal dynamics (blue curves). 
The expected scaling $\propto1/\kappa$ of the protection mechanism is confirmed in panel (f). 
While this example illustrates that the dissipative protection works in principle, we now apply the same mechanism to a more complicated non-Abelian LGT, where enforcing gauge invariance via noise may prove a considerable advantage in the design of an atomic quantum simulator.

\begin{figure}[t]
\includegraphics[width=0.45\textwidth]{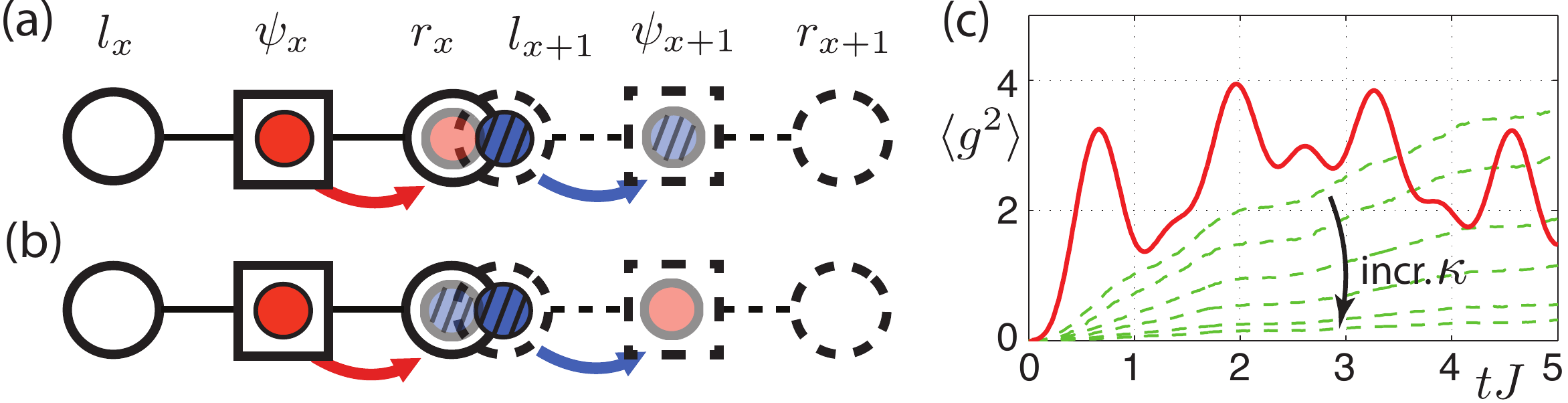}\caption{
U(2) LGT. 
(a) Basic dynamics of Eq.~\eqref{eq:H0_U2}. The matter/gauge-field coupling corresponds to a simultaneous color-conserving tunneling of one fermion at site $x$ to the link $x,x+1$
(red) and a rishon at $x,x+1$ to the site $x+1$ (blue, dashed). Matter and rishon sites are denoted by squares and circles, respectively; the blocks $x$ and $x+1$ are indicated by continuous and dashed contours.
(b) An example for a gauge-variant process: correlated tunneling similar to panel (a), but accompanied by a change of color.
(c) Numerical analysis of two building blocks, evolving under $H_0$ plus various error terms (calculations for the full master equation \eqref{eq:ME}; for details see \cite{supp_dissipative-protection}). 
Red line: $\kappa=0$, green dashed lines: $\kappa/J=5,10,20,40,80,160$ (increasing along the arrow).}
\label{fig:3U2}%
\end{figure}

\emph{Dissipative protection in non-Abelian LGTs.} We now illustrate the dissipative protection of gauge invariance for a non-Abelian LGT, namely a U(2) QLM that may be realistically realized in cold-atom experiments (see below). 
The presence of color (gauge) degrees of freedom allows to investigate in this simplified model physical phenomena related to general non-Abelian gauge theories like QCD, such as chiral symmetry breaking and confinement, and its phase diagram may support exotic condensate phases~\cite{Banerjee2013}.
Its Hamiltonian, which belongs to a class of more general QLMs including U(N) and SU(N) symmetries,
reads $H_{0}=H_{J}+H_{m}$ \cite{footnoteg0}, where $H_{m}=m\sum_{x}(-1)^{x}\psi_{x}^{\alpha\dagger}\psi_{x}^{\alpha}$ describes staggered fermions and $H_J$ is the interaction between matter and gauge field
\begin{equation}
H_{J}=J\sum_{x}\psi_{x}^{\alpha\dagger}\mathcal{U}_{x,x+1}^{\alpha\beta}%
\psi_{x+1}^{\beta}+\text{h.c.}\equiv J\sum_{x}\psi_{x}^{\alpha\dagger}%
r_{x}^{\alpha}l_{x+1}^{\beta\dagger}\psi_{x+1}^{\beta}+\text{h.c.}
\label{eq:H0_U2}%
\end{equation} 
Here, $\alpha,\beta=1,2$ represents the U(2) color degree of freedom (repeated indices are contracted). 
As before, the fermionic matter fields $\psi_{x}^{\alpha}$ live on the vertices of a lattice, 
while for non-Abelian gauge fields it is convenient to represent link variables by \textit{rishon}
fermionic fields $l_{x}^{\alpha}$ and $r_{x}^{\alpha}$ living on the links to the left and right of a given site,  $\mathcal{U}_{x,x+1}^{\alpha\beta}\equiv r_{x}^{\alpha}l_{x+1}^{\beta\dagger}$~\cite{Brower1999,Wiese2013} (see Fig.~\ref{fig:3U2}(a)). 
The U(2) gauge symmetry is split into a U(1) part, with generator
\[
G_{x}=\psi_{x}^{\alpha\dagger}\psi_{x}^{\alpha}-(l_{x+1}^{\alpha\dagger}l_{x+1}^{\alpha}-r_{x}^{\alpha\dagger}r_{x}^{\alpha})/2+(l_{x}^{\alpha\dagger
}l_{x}^{\alpha}-r_{x-1}^{\alpha\dagger}r_{x-1}^{\alpha})/2-1\,,
\]
and a SU(2) part, with generators
\[
G_{x}^{a}=\psi_{x}^{\alpha\dag}\sigma_{\alpha\beta}^{a}\psi_{x}^{\beta}%
+r_{x}^{\alpha\dag}\sigma_{\alpha\beta}^{a}r_{x}^{\beta}+l_{x}^{\alpha\dag
}\sigma_{\alpha\beta}^{a}l_{x}^{\beta}\,,\;\;a=1,2,3.
\]
The $G_{x}^{a}$ commute with all $G_{x}$ and satisfy $[G_{x}^{a},G_{y}^{b}]=2i\delta_{xy}\epsilon
_{abc}G_{x}^{c}$, where $\epsilon_{abc}$ is the Levi--Civita tensor and $\sigma^a$ are Pauli matrices.

The basic system dynamics described by $H_J$ is illustrated in Fig.~\ref{fig:3U2}(a): it corresponds to a simultaneous hopping of two particles, namely $\psi_x$ to $r_x$ and simultaneously $l_{x+1}$ to $\psi_{x+1}$. 
The color of both particles is preserved during the process ($[H_0,G^a_x]=0$). 
In a typical microscopic implementation, one may obtain additional, undesired color-changing terms of the form $H_1^{(1)}= \lambda \sum_x (\psi_{x+1}^{2\dag}l_{x+1}^{1}\,r_{x}^{1\dag}\psi_{x}^{2} + {\rm h.c.})$, as illustrated in Fig.~\ref{fig:3U2}(b). 
These do not commute with all generators, and therefore violate gauge invariance. 
To estimate the effect of terms such as this one, we analyzed the exact time evolution for two building blocks of the U(2) model. 
We included various realistic errors similar to $H_1^{(1)}$ that are specific for the cold-atom implementation described in \cite{supp_dissipative-protection}, comprising a large class of generic errors. 
As Fig.~\ref{fig:3U2}(c) shows, without protection the mean value of the sum of all generators, $g^{2}=\sum_{x} (G_x^2+\sum_a (G^{a}_{x})^{2})/N_{s}$, quickly acquires large values, indicating the loss of gauge-invariance (red line). 
However, under the noise protection generated by Eq.~\eqref{eq:Hclassicalnoise}, gauge invariance may be retained on the time scale of several tunneling events (green dashed lines). This example demonstrates that the proposed protection mechanism works also for more complicated non-Abelian models including several non-commuting generators.

\emph{Optical-lattice implementation.} In ultracold-atom implementations where the color index is represented by different internal atomic states, 
the standard strategy to suppress gauge-variant terms via quadratic energy penalties $U(G_{x}^{a})^{2}$ amounts to engineering numerous local and non-local interactions with \emph{fine-tuned} coefficients. 
From this regard, our dissipative approach is advantageous, since the preservation of gauge invariance requires driving the system with terms that are only linear in the generators. 
In the ultracold-atoms setting,  the on-site \emph{single-particle} noise terms $\xi_{x}^{a}(t)G_{x}^{a}$ can be realized by coupling internal atomic states to laser fields with suitable amplitude or phase noise, where noisy AC-Stark shifts and Raman processes allow to impose the constraints on $G_{x}$ and $G_{x}^{3}$ as well as $G_{x}^{1,2}$ (see \cite{supp_dissipative-protection}). 
Using high-resolution objectives \cite{Bakr2009,Weitenberg2011,Zimmermann2011,Bourgain2012}, it is possible to engineer an independent noise source for each generator, as required by Eq.\,\eqref{eq:Hclassicalnoise}. However, in the common case where the dominant gauge-variant perturbations couple only nearest neighbors, one can simplify the experimental setup by using a noise pattern that is repeated periodically. In this way, one can enforce local gauge invariance by using global addressing together with a superlattice structure \cite{supp_dissipative-protection}.  

The last ingredient to quantum simulate the SU(2) LGT is then a natural realization of $H_0$ that does not interfere with the dissipative protection and thus does not lead to undesired heating of the system by the noise.  
In \cite{supp_dissipative-protection}, we illustrate how models with U(N) interactions (in particular focusing on the conceptually simpler U(2) case) can be engineered in spinor gases, where spin-changing collisions combined with state-dependent optical potentials provide a natural realization of the two-body interaction terms constituting $H_J$. 
Ideas along these lines for Abelian theories have also been discussed in Ref.~\cite{Zohar2013a}.

The scheme outlined above could also be combined with energetic protection in cases where, e.g., the interactions only protect an Abelian symmetry, while the more challenging non-Abelian contributions are imposed via noise. This would facilitate the realization of previous proposals~\cite{Banerjee2013}, extending their regime of applicability and providing additional means to improve the accuracy of gauge invariance in microscopic realizations. 

\emph{Scaling and Imperfections.} 
In contrast to quantum-computing purposes, we are interested here in many-body properties, such as the expectation value of low-order correlations and order parameters~\cite{Hauke2011d}. This ensures, in general, better scalability properties: while the leakage out of the $\mathcal P$ subspace is expected to increase with the system size, order parameters that quantify gauge invariance, such as $g^2$, are not severely affected by the system size itself. 
While checking these expectations for sufficiently large system sizes with LGTs is outside of computational capabilities, we have tested these scalings in the context of a simplified model where local conservation laws are imposed in the same dissipative manner. 
The results are described in Ref.~\cite{supp_dissipative-protection} and clearly support these claims. 

To further address the feasibility of our proposal, we have performed a numerical analysis of typical error sources in realistic setups, such as particle loss and imperfect noise addressing. In particular, we found that
the effects of the latter commonly scale as $\epsilon^2 \kappa$ for short timescales, where $\epsilon$
is the strength of the imperfections, and not as $\epsilon\kappa$ as naively expected. This is due to the particular characteristics of the most common addressing errors, which do not directly affect the gauge invariant subspace (see \cite{supp_dissipative-protection} for details).

\emph{Conclusions and Outlook.} We have shown how classical noise can serve as a resource to engineer constrained Hamiltonian dynamics in quantum simulators, and in particular how Abelian and non-Abelian gauge invariance can
be protected in atomic lattice implementations. The dissipative scheme has
advantages with respect to the more conventional energy punishment, as
coupling to generators is linear, local, and introduced by a physical resource
which is independent of the engineered Hamiltonian dynamics. 
For gauge-variant perturbations that do not couple distant sites, the noise protection can be realized by global beams in a superlattice configuration.
The mechanism is universal, as it can be extended to any symmetry and dimensionality, and can be applied to different 
microscopic systems beyond cold atom gases, such as superconducting qubits and trapped ions~\cite{Marcos2013,Hauke2013b}.

\emph{Acknowledgments.}
We thank D.\ Banerjee, Ch.\ Becker, E.\ Rico, and U.-J.\ Wiese for stimulating discussions and G.\ Kurizki, S.\ Pascazio, and L.\ Viola for useful comments on the manuscript. 
This project was supported by SIQS, the ERC Synergy Grant UQUAM, the SFB FoQuS (FWF Project No. F4006-N16), U.S.\ ARO MURI award W911NF0910406,  and the START grant Y 581-N16 (SD).
M.H.\ thanks IQOQI, Innsbruck for hospitality.

\renewcommand{\theequation}{S\arabic{equation}}
\setcounter{equation}{0}
\renewcommand{\thefigure}{S\arabic{figure}}
\setcounter{figure}{0}

\newpage
\onecolumngrid

{
\center \bf \Large 
Supplemental Material to\vspace*{0.1cm}\\ 
\emph{Constrained dynamics via the Zeno effect in quantum simulation:\vspace*{0.1cm}\\ 
\hspace*{0.7cm}Implementing non-Abelian lattice gauge theories with cold atoms}
}

\vspace*{0.5cm}
{\center{
\hspace*{0.1\columnwidth}\begin{minipage}[c]{0.8\columnwidth}
In this supplemental material, we show how classical noise leads to an effective evolution within a gauge-invariant subspace, suppressing gauge-variant errors. 
We also present a possible implementation of a non-Abelian lattice gauge theory that does not interfere with the engineered noise, relying on spin-changing scattering processes between two fermionic species in a spin-dependent optical lattice. Moreover, we show on a simplified model that local observables are protected independently of system size. 
\end{minipage}
}
}

\section{Noise Protection in the Master Equation Formulation}

Here, we present the derivation of the effective master equation described by Eq.~(3) of the main text. This provides an intuitive explanation of the non-Hermitian part of the effective Hamiltonian $H_{\rm eff}$ in terms of a loss of population to the gauge-variant subspace.

\subsection{Elimination of classical noise variables} 
\label{sec:noise}

We start from the Hamiltonian coupled to classical Gaussian white noise variables, Eq.~(1) in the main text.  
The associated time evolution  of the system density matrix is governed by the master equation, \cite{Carmichael1993sup,Gardiner2000sup}
\begin{align}
\label{eq:MEnoise}
\dot\rho&= \mL_0 \rho +\sum_{x,a} \xi_x^a(t)\mL_x^a\rho ,\\
\mL_0\rho&=-i[ H,\rho], \quad 
\mL_x^a\rho=-i\sqrt{2\kappa}[G_x^a,\rho].
\end{align}
Here, the total Hamiltonian is given as $H = H_0 + H_1$ with $H_0 (H_1)$ the gauge invariant (gauge variant) pieces as in the main text. We interpret \eeqref{eq:MEnoise}  as a Stratonovich stochastic differential equation (SDE) $d\rho=\mL_0 dt +\sum_{x,a} dW_x^a \mL_x^a\rho $, where the $dW_x^a$ are independent Wiener increments \cite{Gardiner2004sup}. This equation can be converted to an Ito SDE, which then allows us to average over the classical noise in a straightforward manner. As a result of this procedure, we obtain the following master equation for the noise-averaged density operator $\lat\rho\rat$ :
\begin{align}
\lat\dot \rho\rat
&= \mL_0 \lat\rho\rat + \tfrac{1}{2} \sum_{x,a} \mL_x^a\mL_x^a\lat\rho\rat = -i[H,\lat \rho\rat ] + \kappa \sum_{x,a}  ( 2 G_x^a \lat\rho\rat G_x^a -  (G_x^a)^2 \lat \rho\rat - \lat\rho\rat (G_x^a)^2 )\, .\label{eq:gn}
\end{align}
This is the result quoted in Eq.~(2) of the main text, where we dropped the angular brackets for notational simplicity. Note that $\kappa$ can be much larger than the system energy scales entering $H_0,H_1$, respectively.

\subsection{Projected Master Equation}

We are interested in the detrimental situation where the scales in the total Hamiltonian $H = H_0 + H_1$, $H_0 \sim J$ and $H_1\sim \lambda$, are of the same order, $J\approx \lambda$. On the other hand, the scale of the dissipative terms $\sim \kappa$ is assumed to obey $\kappa \gg \lambda$. 
In this regime, we can adiabatically eliminate the fast dynamics generated by the dissipative part of the evolution. 
Note that, crucially, the desired gauge-invariant Hamiltonian $H_0$ and the gauge-variant perturbation $H_1$ can involve comparable energy scales; the protection mechanism builds only on the noise level $\kappa$ being the largest scale in the problem.  
We introduce projectors $\mathcal P$ and $\mathcal Q = 1-\mathcal P$, where $\mathcal P$ projects on the gauge invariant subspace $\mH_\mP$ of the Hilbert space, as in the main text. The operators read in these projections (we replace $\lat \rho\rat \to \rho$ for notational simplicity),   
\begin{equation}
\rho  = \left( \begin{array}{cc}
\rho_{\mathcal{PP}} &  \rho_{\mathcal{PQ}}\ \\
 \rho_{\mathcal{QP}} &  \rho_{\mathcal{QQ}}  
\end{array}
\right),\quad 
H = \left( \begin{array}{cc}
 H_{\mathcal{PP}} & H_{\mathcal{PQ}}\ \\
 H_{\mathcal{QP}} &  H_{\mathcal{QQ}}  
\end{array}
\right),\quad 
G_x^a =  (G_x^a)^\dag = \left( \begin{array}{cc}
0 &0 \\
0 & G^a_{x}  
\end{array}
\right),
\end{equation}
where $\rho_{\mathcal{PP}} = \mathcal P \rho \mathcal P$ etc. In particular, the Hermitian gauge generators act nontrivially only in the gauge-variant subspace. Adiabatically eliminating the coherences $\rho_{\mathcal{PQ}}$, we find the evolution in the physical subspace
\begin{eqnarray}\label{eq:effmotion}
\partial_t \rho_{\mathcal{PP}} \approx -i [H_{\mathcal{PP}},\rho_{\mathcal{PP}}]  - H_{\mathcal{PQ}}(\kappa  \sum_{x,a} G^a_{x} G^a_{x} )^{-1}_{\mathcal{QQ}}H_{\mathcal{QP}}\rho_{\mathcal{PP}}  - \rho_{\mathcal{PP}} H_{\mathcal{PQ}}  ( \kappa \sum_{x,a} G^a_{x} G^a_{x} )_{\mathcal{QQ}}^{-1} H_{\mathcal{QP}} , 
\end{eqnarray}
which is the master equation for the effective Hamiltonian given in Eq.~(3) of the main text 
(note for comparison with the text $H_{\mathcal{PQ}} = \mathcal P H \mathcal Q = \mathcal P H_1 \mathcal Q $ and analogous for $H_{\mathcal{QP}}$; $H_0$ has vanishing matrix elements in the above projection). 
This equation does not have Lindblad form and therefore may seem to violate the conservation of probability. To see that this is in fact not the case, we have to study $\partial_t \mathrm{tr} \rho  = \partial_t \mathrm{tr} \rho_{\mathcal{PP}} + \partial_t \mathrm{tr} \rho_{\mathcal QQ}$, where $\partial_t \rho_{\mathcal QQ}= -i \big(  H_{\mathcal{QP}}\rho_{\mathcal{PQ}} - \rho_{\mathcal{QP}}H_{\mathcal{PQ}}\big)  +\kappa \sum_{x,a}[ 2 G^a_{x}\rho_{\mathcal QQ} G^a_{x} -  \{ G^a_{x}G^a_{x}, \rho_{\mathcal QQ}\}]$ in the above decomposition. There is an explicit cancellation between the terms, which separately do not preserve the norm. 
The net effect of the second and third terms in \eqref{eq:effmotion}, therefore, is a loss of probability to be in the gauge-invariant subspace.

\section{Spin-changing collisions and local noise operations for optical lattice implementations}

In this section, we explain how to implement the two crucial ingredients for a quantum simulation of a U(2) lattice gauge theory, (i) a way to generate the dynamics $H_0$ that is robust against the added noise, and (ii) a possibility to couple the noise to the symmetry generators. 

\subsection{Correlated hopping from spin-changing collisions}

In the rishon representation as used in Eq.~(5) of the main text, the correlated hopping term $H_J$ is a product of four field operators. 
In a gas of ultracold atoms, such products appear naturally in the form of atom--atom scattering processes, providing us with an elegant possibility of realizing $H_0$ via spin-changing collisions. 

\subsubsection{Scattering Hamiltonian}

In second quantization, two-body collisions between atoms are described by the Hamiltonian 
\begin{equation}
\hHint = \frac 1 2 \sum_{\alpha,\beta,\gamma,\delta} \int \rmd^3r\rmd^3r^\prime \Oopd_\gamma(\vect{r}) \Oopd_\delta(\vect{r}^\prime) \hV_{\alpha,\beta}^{\gamma,\delta}(\vect{r},\vect{r}^\prime) \Oop_\beta(\vect{r}^\prime) \Oop_\alpha(\vect{r})\,. 
\end{equation}
Here, the $\Oop_\alpha$ denote (fermionic) field operators and the greek indices $\alpha\equiv\left\{s_\alpha,m_\alpha\right\}$ summarize internal states $m_\alpha$ as well as a species index $s_\alpha$. 
At low temperatures, we can express the interaction matrix elements as a contact interaction $\hV_{\alpha,\beta}^{\gamma,\delta}(\vect{r},\vect{r}^\prime)=V_{\alpha,\beta}^{\gamma,\delta}\delta(\vect{r}-\vect{r}^\prime)$, which gives
\begin{equation}
\label{eq:HintContact}
\hHint = \frac 1 2 \sum_{\alpha,\beta,\gamma,\delta} \int \rmd^3r\Oopd_\gamma(\vect{r}) \Oopd_\delta(\vect{r})\, V_{\alpha,\beta}^{\gamma,\delta}\,\Oop_\beta(\vect{r}) \Oop_\alpha(\vect{r})\,.
\end{equation}
The interaction obeys the Pauli exclusion principle, it conserves angular momentum, $m_{\delta}=m_{\alpha}+m_{\beta}-m_{\gamma}$, and it cannot change the species, $s_\alpha=s_\delta$ and $s_\beta=s_\gamma$. 
Moreover, for fermions, the interaction matrix elements are anti-symmetric $V_{\beta,\alpha}^{\gamma,\delta}=V_{\alpha,\beta}^{\delta,\gamma}=-V_{\alpha,\beta}^{\gamma,\delta}$ etc.

The $V_{m_1,m_2}^{m_1^\prime,m_2^\prime}$ (where we suppressed the species index) are not all independent, but can be expressed by just a few parameters $V_F$, where $F$ denotes the total spin of the two-particle eigenstates, with associated magnetic quantum number $M_F$. 
For atoms with individual spins $F_{1,2}$, one has \cite{Krauser2012sup}
\begin{equation}
 	V_{m_1,m_2}^{m_1^\prime,m_2^\prime} = \sum_{F} \braket{F_1,m_1^\prime,F_2,m_2^\prime|F,M_F} \braket{F,M_F|F_1,m_1,F_2,m_2} V_{F} \,, 
\end{equation}
where the $\braket{F_1,m_1,F_2,m_2|F,M_F}$ are Clebsch--Gordan coefficients. 
For homonuclear scattering, symmetry requirements restrict the total spin to $F=0,2,\dots,F_1+F_2-2,F_1+F_2$, whereas for heteronuclear scattering one has $F=0,1,\dots,F_1+F_2-1,F_1+F_2$. 

\begin{figure}
\centering
\includegraphics[width=0.85\columnwidth]{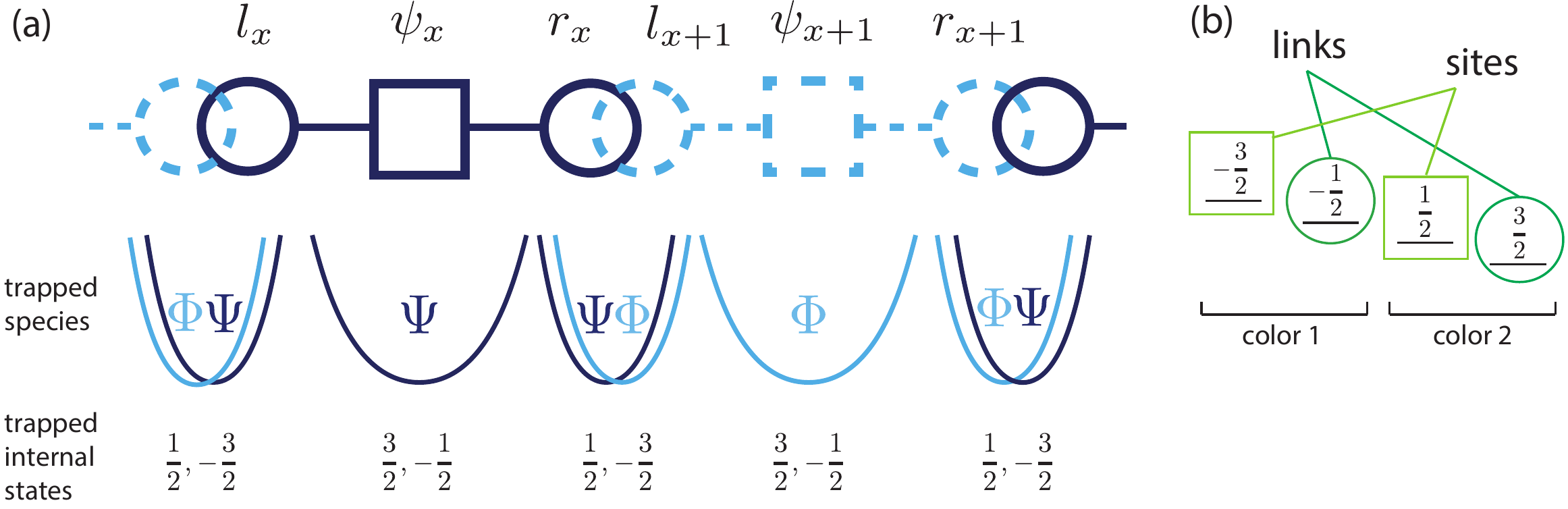}
\caption{Implementation of the non-Abelian model. 
(a) The U(2) model sketched on the top can be realized by the arrangement of the spin-dependent optical lattice given at the bottom. 
$\Phi$ and $\Psi$ are two fermionic atom species with internal states $m=-\frac 3 2, -\frac 1 2, \frac 1 2, \frac 3 2$, trapped as indicated. 
(b) Mapping of internal states to colors.
}
\label{fig:implementation}
\end{figure}

\subsubsection{Expansion in Wannier functions of the optical lattice}

The main idea to pass from the scattering Hamiltonian \eqref{eq:HintContact} to $H_J$ is to associate a spatial dependence to certain scattering processes. This can be achieved via a spin-dependent optical lattice. 
Since in that case different spin states occupy different lattice wells, certain spin-changing collisions are necessarily accompanied by hopping processes (similar to the Abelian implementation proposed in \cite{Zohar2013asup}). 
As a concrete example, we consider two fermionic species, denoted by $s=\Psi,\Phi$, each with four internal degrees of freedom $m=-\frac 3 2,-\frac 1 2, \frac 1 2, \frac 3 2$ (for simplicity, we assume that these are the sublevels of a $F_{1,2}=3/2$ hyperfine manifold). 
We denote the associated field operators by $\Oop_{\Phi,m_\alpha}(\vect{r})\equiv\Phi_{m_\alpha}(\vect{r})$ and $\Oop_{\Psi,m_\alpha}(\vect{r})\equiv\Psi_{m_\alpha}(\vect{r})$.
The atoms are loaded in the species-dependent optical lattice sketched in \figref{fig:implementation}(a).
Every second well will represent a site $x$, and can only be populated by a single species in an alternating fashion; e.g., $\Phi$ for $x$ even and $\Psi$ for $x$ odd.
Further, the sites trap only two internal states, say, $m=-1/2,3/2$. 
The links $\braket{x,x+1}$ are represented by the remaining wells. They can be populated by both species but only by the remaining two internal states ($m=-3/2,1/2$). 
Spin-dependent optical lattices (although in simpler configurations) are routinely produced in experiment \cite{Mandel2003sup,Lee2007sup,McKay2010sup,SoltanPanahi2011sup}. 
To illustrate the main idea how to obtain $H_J$ from spin-changing collisions, we consider in the following first an idealized case. Afterwards, we show that in more realistical situations various imperfections appear. These, however, can be sufficiently suppressed by the engineered noise. 

In a sufficiently deep optical lattice, we can as usual expand the field operators in terms of Wannier functions, 
\bal
\label{eq:operatorWannierExpansion}
\Oop_\alpha(\vect{r})=\sum_{i\in \mA_\alpha} \Oop_{i,\alpha} w_i(\vect{r})\,,
\eal
where $\Oop_{i,\alpha}$ annihilates a fermion in state $\alpha$ at well $i$, with associated Wannier function $w_i(\vect{r})$. 
The sum runs over the set of wells $\mA_\alpha$ that are accessible to atoms in state $\alpha$. 
In the following, odd wells become associated to sites $x$ of the quantum link model by $(i+1)/2 = x$, and even lattice wells denote the corresponding links $\braket{x,x+1}$. 
To achieve the desired correlated hoppings, we choose a spin-dependent lattice where 
$\mA_\alpha=\left\{x\, |\, x\, \mathrm{even}\right\}$ if $s_\alpha=\Phi$ and $m_\alpha=-1/2,3/2$, 
$\mA_\alpha=\left\{x\, |\, x\, \mathrm{odd}\right\}$ if $s_\alpha=\Psi$ and $m_\alpha=-1/2,3/2$, and
$\mA_\alpha=\left\{\braket{x,x+1}\right\}$ if $m_\alpha=-3/2,1/2$. 
For simplicity, we assume that $w_i(\vect{r})$ depends only on whether the well denotes a link or a site, but not on species or spin state. 

In this Wannier expansion (using, without loss of generality, real Wannier functions), the collision Hamiltonian becomes 
\begin{equation}
\label{eq:Hint_Wannier}
\hHint = \frac 1 2 \sum_{\alpha,\beta,\gamma,\delta}\,\, \sum_{i\in \mA_\alpha} \sum_{j\in \mA_\beta} \sum_{k\in \mA_\gamma} \sum_{\ell\in \mA_\delta} 
V_{\alpha,\beta}^{\gamma,\delta} \int \rmd^3r\, w_i(\vect{r}) w_j(\vect{r}) w_k(\vect{r}) w_\ell(\vect{r})
\Oopd_{k,\gamma} \Oopd_{\ell,\delta} \Oop_{j,\beta} \Oop_{i,\alpha}\,. 
\end{equation}
Already for moderately deep lattices, due to the exponential decay of Wannier functions, we can neglect terms involving wells that are not nearest neighbors or constituted by the triples $x-1,\braket{x-1,x},x$ (we assume that the wells associated to the links $\braket{x,x+1}$ are much deeper than wells associated to the sites $x$, so that the overlap of triples $\braket{x-1,x},x,\braket{x,x+1}$ can be neglected).  

Due to the spin-dependence of the optical lattice, collision terms including the three wells $x-1,\braket{x-1,x},x$ are necessarily spin-changing (see~\figref{fig:implementation}(a)). 
If, for the moment, we assume that all corresponding scattering rates are equal, these collision terms yield the Hamiltonian 
\begin{align}
\label{eq:scat1}
H_{{\rm int}}^{\prime}/J&=\sum_{x\,\mathrm{even}} \sum_{\sigma,\sigma^\prime=\frac 3 2,-\frac 1 2} \Bigl(
\Phi_{x}^{\sigma\,\, \dag } \Phi_{x,x+1}^{\sigma-1} 
\Psi_{x,x+1}^{\sigma^{\prime}-1\, \dag}  \Psi_{x+1}^{\sigma^\prime}  +\hc \Bigr) + \sum_{x\,\mathrm{odd}} \Phi\leftrightarrow\Psi \\
&+ \sum_{x\,\mathrm{even}} \Bigl[
\Phi_{x,x+1}^{\frac 1 2 \,\, \dag } \Phi_{x}^{-\frac 1 2 } 
\Psi_{x+1}^{-\frac 1 2 \, \dag}  \Psi_{x,x+1}^{\frac 1 2 } 
+
\hspace*{-6pt}
\sum_{\sigma=\frac 3 2,-\frac 1 2} 
\Bigl(
\Phi_{x}^{\sigma\,\, \dag } \Phi_{x,x+1}^{\sigma-1} 
\Psi_{x+1}^{-\frac 1 2 \, \dag}  \Psi_{x,x+1}^{\frac 1 2 } 
+ \Phi_{x,x+1}^{\frac 1 2 \,\, \dag } \Phi_{x}^{-\frac 1 2 } 
\Psi_{x,x+1}^{\sigma-1\, \dag}  \Psi_{x+1}^{\sigma}\Bigr) 
+\hc
\Bigr] 
+ \sum_{x\,\mathrm{odd}} \Phi\leftrightarrow\Psi\,, \nonumber
\end{align}
where $J\equiv J_{\alpha,\beta}^{\gamma,\delta}=V_{\alpha,\beta}^{\gamma,\delta}\int \rmd^3r\, w_{x}(\vect{r}) w_{x,x+1}^2(\vect{r}) w_{x+1}(\vect{r}) $, with $s_\alpha=s_\delta=\Phi$ and $s_\beta=s_\gamma=\Psi$. 

\begin{figure}
\centering
\includegraphics[width=0.8\columnwidth]{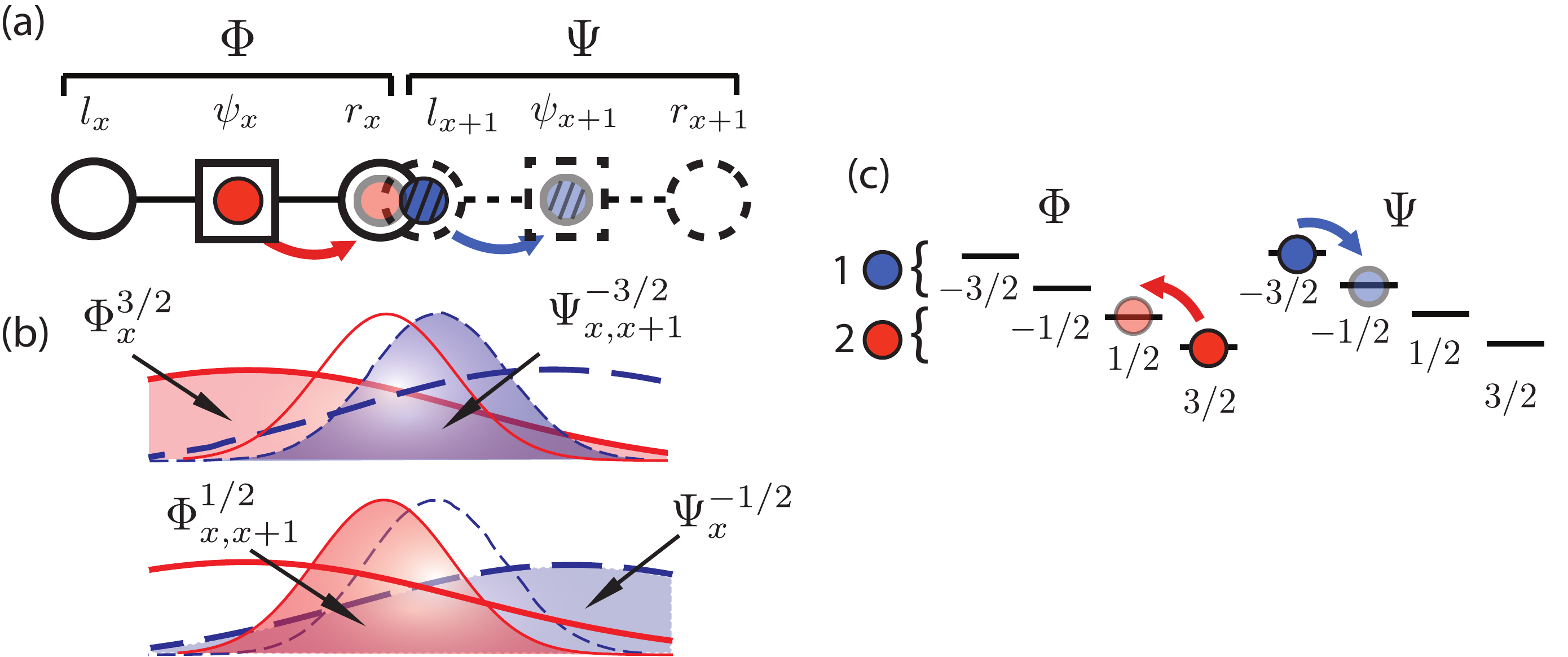}
\caption{Microscopic process generating the correlated hopping $H_J$; 
(a,b) real-space and (c) spin-space representation. 
(a) The correlated hopping corresponds to a fermion of one color (red) jumping to an adjacent rishon site with a simultaneous hopping of a rishon (sketched here for a different color, blue) to the next site. 
(b) Occupied Wannier functions before (top) and after the hopping (bottom) are shaded. The correlated hopping is generated by a spin-changing collision taking place in the region where all four Wannier functions overlap. 
(c) The collision changes the internal states of a fermion of species $\Phi$ and one of $\Psi$, which has to be accompanied by a hopping process due to the spin dependence of the optical lattice (as sketched in Fig.~\ref{fig:implementation}(a)). 
}
\label{fig:correlated_tunneling}
\end{figure}
The first line of Eq.~\eqref{eq:scat1} describes simultaneous tunneling of one particle from a site $x+1$ to a link $\braket{x,x+1}$ and of a second particle from the link to the neighboring site $x$. 
This gives exactly the desired correlated tunneling terms $H_J$ if we use the mapping 
\begin{subequations}
\label{eq:mapping}
\begin{align}
&\hspace*{-0.5cm}\mathrm{for\,\, even}\,\, x:\nonumber
\\
&\Phi_{x}^{3/2,-1/2}\longrightarrow \psi_{x}^{1,2}\,,\quad
\Phi_{x,x+1}^{1/2,-3/2}\longrightarrow r_{x}^{1,2}\,,\quad
\Phi_{x-1,x}^{1/2,-3/2}\longrightarrow l_{x}^{1,2}\,,\quad
\Psi_{x,x+1}^{1/2,-3/2}\longrightarrow l_{x+1}^{1,2}\,,\quad
\Psi_{x-1,x}^{1/2,-3/2}\longrightarrow r_{x-1}^{1,2}\,, 
\\
&\hspace*{-0.5cm}\mathrm{for\,\, odd}\,\, x:\nonumber
\\
&\Psi_{x}^{3/2,-1/2}\longrightarrow \psi_{x}^{1,2}\,,\quad
\Psi_{x,x+1}^{1/2,-3/2}\longrightarrow r_{x}^{1,2}\,,\quad
\Psi_{x-1,x}^{1/2,-3/2}\longrightarrow l_{x}^{1,2}\,,\quad
\Phi_{x,x+1}^{1/2,-3/2}\longrightarrow l_{x+1}^{1,2}\,,\quad
\Phi_{x-1,x}^{1/2,-3/2}\longrightarrow r_{x-1}^{1,2}\,.
\end{align}
\end{subequations}
In the quantum link model, we interpret the operators $\psi_x^{1,2}$ as fermions on sites $x$, and $r_x^{1,2}$ and $l_x^{1,2}$ as the associated left and right rishons. All of these appear in two colors, labeled 1 and 2. Their translation back to the physical internal states can be found in \figref{fig:implementation}(b).
Via this mapping, the first line of Eq.~\eqref{eq:scat1}, gives rise to the desired dynamics 
\begin{subequations}
\label{eq:Hintideal}
\begin{align}
H_{J}=&J\sum_x\left[  
  \psi_{x-1}^{2\dag} r_{x-1}^{2} \,l_{x}^{1\dag} \psi_x^{1} 
+ \psi_{x-1}^{1\dag} r_{x-1}^{1} \,l_{x}^{1\dag} \psi_x^{1}
+ \psi_{x-1}^{2\dag} r_{x-1}^{2} \,l_{x}^{2\dag} \psi_x^{2}
+ \psi_{x-1}^{1\dag} r_{x-1}^{1} \,l_{x}^{2\dag} \psi_x^{2} \right]+\hc\\
=&J\sum_x\sum_{c,d=1,2}\psi_{x-1}^{c\dag} r_{x-1}^c l_{x}^{d\dag} \psi_{x}^d +\hc 
\end{align}
\end{subequations}
The associated process is explained in Fig.~\ref{fig:correlated_tunneling}.

\begin{table*}
\begin{tabular}{ l l c r l c r  }
\toprule
\quad\quad\quad\quad &
process &
species &
\quad\quad $m_\alpha,m_\beta;$ &
$m_\gamma,m_\delta$ \quad\quad&
Wannier \quad &
used value
\tabularnewline
&
&
&
&
&
overlap &
$\left[J\right]$ 
\tabularnewline
\cline{1-7}
\tabularnewline
$J$ &
$\sum_{c,d=1,2}\psi_{x-1}^{c\dag} r_{x-1}^c l_{x}^{d\dag} \psi_{x}^d +\hc$ &
&\hspace*{-3cm}
$J=(J_{11}+J_{12})/2$
&
&
&
1
\vspacefortable 
\vspacefortable 
\tabularnewline
$J_{11}$&
$\sum_{c=1,2} \psi_{x-1}^{c\dag} r_{x-1}^c l_{x}^{c\dag} \psi_{x}^c+\hc$ &
$\Phi\Psi$ &
$\frac 3 2,\frac 1 2;$ &
$\frac 1 2,\frac 3 2$ &
$\mI_{0}$ &
$2.03$
\vspacefortable \tabularnewline
$J_{12}$&
$\sum_{c=1,2}\psi_{x-1}^{c\dag} r_{x-1}^c l_{x}^{\bar c \dag} \psi_{x}^{\bar c}+\hc$ &
$\Phi\Psi$ &
$-\frac 1 2,\frac 1 2;$ &
$-\frac 3 2,\frac 3 2$ &
$\mI_{0}$ &
$-0.03$
\vspacefortable 
\tabularnewline
\tabularnewline
&
$\sum_{c=1,2} \left( \psi_x^{2\dag} l_{x}^{1} \psi_{x-1}^{c\dag}  r_{x-1}^{c}
+ l_{x}^{c\dag} \psi_x^{c} r_{x-1}^{1\dag} \psi_{x-1}^{2} \right)+\hc$ 
&
$\Phi\Psi$ &
$\frac 1 2,-\frac 3 2;$ &
$-\frac 1 2,-\frac 1 2$ &
$\mI_{0}$ &
$0.05$
\vspacefortable \tabularnewline
&
$\psi_x^{2\dag} l_{x}^{1} r_{x-1}^{1\dag} \psi_{x-1}^{2}+\hc$
&
$\Phi\Psi$ &
$\frac 1 2,-\frac 1 2;$ &
$-\frac 1 2,\frac 1 2$ &
$\mI_{0}$ &
$1.97$
\vspacefortable \tabularnewline
&
$\psi_x^{2\dagger}r_x^{2}r_x^{1\dagger}\psi_x^1+\psi_x^{2\dagger}l_x^{2}l_x^{1\dagger}\psi_x^1+\hc$
&
$\Phi\Phi\,,\Psi\Psi$ &
$\frac 3 2,-\frac 3 2;$ &
$\frac 1 2,-\frac 1 2$ &
$\mI_{1}$ &
$2\times 0.32\,\,,\,\,2\times 0.3$
\vspacefortable \tabularnewline
&
$n_{x}^1 n_{x}^2$ &
$\Phi\Phi\,,\Psi\Psi$ &
$\frac 3 2,-\frac 1 2;$ &
$\frac 3 2,-\frac 1 2$ &
$\mI_{2}$ &
$2\times 0.46\,\,,\,\,2\times 0.37$
\vspacefortable \tabularnewline
&
$n_{r,x}^1 n_{r,x}^2+n_{l,x}^1 n_{l,x}^2$ &
$\Phi\Phi\,,\Psi\Psi$ &
$\frac 1 2,-\frac 3 2;$ &
$\frac 1 2,-\frac 3 2$ &
$\mI_{3}$ &
$\cancel{2\times 0.55\,\,,\,\,2\times 0.44}$
\vspacefortable \tabularnewline
&
$n_{r,x}^1n_{l,x+1}^2+n_{r,x}^2n_{l,x+1}^1$ &
$\Phi\Psi$ &
$-\frac 3 2,\frac 1 2;$ &
$-\frac 3 2,\frac 1 2$ &
$\mI_{3}$ &
$\cancel{4.21}$
\vspacefortable \tabularnewline
&
$n_{r,x}^1n_{l,x+1}^1$ &
$\Phi\Psi$ &
$\frac 1 2,\frac 1 2;$ &
$\frac 1 2,\frac 1 2$ &
$\mI_{3}$ &
$\cancel{64.77}$
\vspacefortable \tabularnewline
&
$n_{r,x}^2n_{l,x+1}^2$ &
$\Phi\Psi$ &
$-\frac 3 2,-\frac 3 2;$ &
$-\frac 3 2,-\frac 3 2$ &
$\mI_{3}$ &
$\cancel{66.54}$
\vspacefortable \tabularnewline
&
$\sum_{c=1,2} n_{x}^c (n_{r,x}^c+n_{l,x}^c)$ &
$\Phi\Phi\,,\Psi\Psi$ &
$\frac 3 2,\frac 1 2;$ &
$\frac 3 2,\frac 1 2$ &
$\mI_{1}$ &
$2\times 0.09\,\,,\,\,2\times 0.07$
\vspacefortable \tabularnewline
&
$\sum_{c=1,2} n_{x}^c (n_{r,x}^{\bar c}+n_{l,x}^{\bar c}) $ &
$\Phi\Phi\,,\Psi\Psi$ &
$\frac 3 2,-\frac 3 2;$ &
$\frac 3 2,-\frac 3 2$ &
$\mI_{1}$ &
$2\times 0.42\,\,,\,\,2\times 0.37$
\vspacefortable \tabularnewline
&
$\sum_{c=1,2} n_{x}^c (n_{r,x-1}^c+n_{l,x+1}^c)$ &
$\Phi\Psi$ &
$\frac 3 2,\frac 1 2;$ &
$\frac 3 2,\frac 1 2$ &
$\mI_{1}$ &
$0.92$
\vspacefortable \tabularnewline
&
$n_{x}^1 (n_{r,x-1}^2+n_{l,x+1}^2) $&
$\Phi\Psi$ &
$\frac 3 2,-\frac 3 2;$ &
$\frac 3 2,-\frac 3 2$ &
$\mI_{1}$ &
$0.87$
\vspacefortable \tabularnewline
&
$n_{x}^2 (n_{r,x-1}^1+n_{l,x+1}^1)$ &
$\Phi\Psi$ &
$\frac 1 2,-\frac 1 2;$ &
$\frac 1 2,-\frac 1 2$ &
$\mI_{1}$ &
$1.16$
\vspacefortable \tabularnewline
&
$r_{x}^{2\dagger} r_x^1 l_{x+1}^{1\dagger} l_{x+1}^2 +\hc$ &
$\Phi\Psi$ &
$-\frac 1 2,\frac 3 2;$ &
$\frac 3 2,-\frac 1 2$ &
$\mI_{3}$ &
$\cancel{59.67}$
\vspacefortable \tabularnewline
&
$\psi_{x}^{2\dagger} \psi_x^1 l_{x+1}^{1\dagger} l_{x+1}^2 + \psi_{x}^{2\dagger} \psi_x^1 r_{x-1}^{1\dagger} r_{x-1}^2 +\hc$ &
$\Phi\Psi$ &
$\frac 3 2,-\frac 3 2;$ &
$-\frac 1 2,\frac 1 2$ &
$\mI_{1}$ &
$0.39$
\vspacefortable \tabularnewline
&
$\psi_{x+1}^{1\dagger} l_{x+1}^2 r_{x}^{2\dagger} \psi_{x}^1 +\hc$ &
$\Phi\Psi$ &
$\frac 3 2,-\frac 3 2;$ &
$-\frac 3 2,\frac 3 2$ &
$\mI_{0}$ &
$1.91$
\vspacefortable \tabularnewline
\botrule
\end{tabular}
\caption{
Interaction strengths for the scattering processes contained in $H_{\rm int}$. 
For reference, the first row contains the desired correlated hopping $H_J$, which is generated by the physical processes in the second and third row. 
The difference between $J_{11}$ and $J_{12}$ leads to the gauge-variant perturbation given in Eq.~\eqref{eq:deltaJcd}. 
All other scattering processes of $H_{\rm int}$ constitute additional error terms. 
The second column gives the processes in terms of the rishon representation of the U(2) model. 
Here, we defined $\bar c=(c \mod 2)+1$, as well as $n_{x}^c=\psi_x^{c\dagger} \psi_x^c$, $n_{r,x}^c=r_x^{c\dagger} r_x^c$, and $n_{l,x}^c=l_x^{c\dagger} l_x^c$. 
The second (third) column contains the involved species (internal levels). 
The fourth column gives the Wannier overlaps entering the interaction strengths, which read 
$\mI_{0}=\int \rmd^3r\, w_{x-1,x}(\vect{r}) w_{x}^2(\vect{r}) w_{x,x+1}(\vect{r})$, 
$\mI_{1}=\int \rmd^3r\, w_{x}^2(\vect{r}) w_{x,x+1}^2(\vect{r})$, 
$\mI_{2}=\int \rmd^3r\, w_{x}^4(\vect{r})$, and 
$\mI_{3}=\int \rmd^3r\, w_{x,x+1}^4(\vect{r})$. 
Due to the different extension of Wannier functions (wells associated to links are deeper), their strength will increase from $\mI_{0}$ to $\mI_{3}$, with precise values depending on the implementation. In our numerical calculations, we choose the relative weights $1:5:25:30$. 
Furthermore, we use the scattering lengths (with $a_0$ the Bohr radius) 
$V_{F=0}^{\Psi\Phi} = -220 a_0$, 
$V_{F=1}^{\Psi\Phi} = 280 a_0$, 
$V_{F=2}^{\Psi\Phi} = -250 a_0$, 
$V_{F=3}^{\Psi\Phi} = 300 a_0$, 
as well as 
$V_{F=0}^{\Phi\Phi} = 36 a_0$, 
$V_{F=2}^{\Phi\Phi} = 4 a_0$, 
and
$V_{F=0}^{\Phi\Phi} = 40 a_0$, 
$V_{F=2}^{\Phi\Phi} = 5 a_0$. 
The actual values of these parameters will depend on details of the experimental realization. They are chosen here to give a sensible order-of-magnitude estimate. We adjusted them only in a rough way in order to keep the most detrimental errors small, but assumed no fine tuning. 
From these scattering lengths, together with the Wannier overlaps, one obtains the relative interaction strengths as given in the last column. 
In cases where there are two entries, the first one is for even and the second for odd sites (corresponding to intra-species scattering of $\Phi$ and $\Psi$, respectively). 
Intra-species scattering processes have a factor of 2 with respect to inter-species collisions due the contribution from exchange interactions. 
	Finally, in our numerical calculations, we assume a representation where there is only one fermionic rishon per link, such that the density-density interactions which are barred out do not contribute (see text).
\label{tab:scatteringMatrixElements}
}
\end{table*}

Besides the desired terms, the interaction Hamiltonian \eqref{eq:Hint_Wannier} leads to a series of undesired terms, which can reach energy scales comparable to the desired ones. 
For example, the second line of Eq.~(\ref{eq:scat1}) contains color-changing processes that are gauge-variant perturbations. 
Following the mapping \eqref{eq:mapping}, these can be written as 
\begin{equation}
H_1^{(1)}=\psi_x^{2\dag} l_{x}^{1} r_{x-1}^{1\dag} \psi_{x-1}^{2} + \sum_{c=1,2} \left( \psi_x^{2\dag} l_{x}^{1} \psi_{x-1}^{c\dag}  r_{x-1}^{c}
+ l_{x}^{c\dag} \psi_x^{c} r_{x-1}^{1\dag} \psi_{x-1}^{2} \right)+\hc\,.
\end{equation}
Another perturbation appears when we relax the idealized assumptions leading to Eq.~\eqref{eq:scat1}. Without unrealistic fine tuning, the scattering elements $J_{\alpha,\beta}^{\gamma,\delta}$ will not be all equal because of differences in the Clebsch--Gordan coefficients for different spin states. 
One can capture this effect by including the error term
\begin{align}
\label{eq:deltaJcd}
H_1^{(2)}=\sum_x
\sum_{c,d=1,2}
\Delta{J}_{cd}\,\, \psi_{x-1}^{c\dag} r_{x-1}^c l_{x}^{d\dag} \psi_{x}^d +\hc\,. 
\end{align}
In the formulation of this error term, we exploited symmetries of the microscopic scattering interaction to reduce the number of independent scattering matrix elements to just two, namely $J_{12}= J_{21}$ and $ J_{11}= J_{22}$. These determine the strength of the desired correlated hopping, $J\equiv (J_{11}+J_{12})/2$, and of the error term $\Delta{J}_{cd}\equiv {J}_{cd}-J$. 
Other error terms appear due to scattering processes that do not involve all three wells of a building block, such as on-site density--density interactions. 
All of these terms are listed in Table~\ref{tab:scatteringMatrixElements}, along with exemplary values for their strength, which are used in the numerics (see below). Besides the error terms appearing due to scattering processes, as listed in Table~\ref{tab:scatteringMatrixElements}, gauge variance could be broken by single-particle tunneling. 
In the above geometry, however, the Wannier functions of the rishon wells are strongly localized, so that rishon tunneling can be neglected. 
Additionally, the sites are occupied by two species in an alternating pattern, precluding nearest-neighbor tunneling between sites $x$ and $x+1$. 
Therefore, single-particle tunneling is strongly suppressed compared to the desired dynamics and other error sources.

\subsubsection{Numerical simulation of a small system}

To illustrate the dissipative protection scheme for the case of the U(2) quantum link model, we have performed exact numerical simulations of a small system consisting of two building blocks with periodic boundary conditions. The results are displayed in Fig.~3(b) of the main text and were obtained by integrating the master equation (2) of the main text, with the Hamiltonian containing all terms listed in Table~\ref{tab:scatteringMatrixElements}. The gauge-invariant initial state was chosen to be
\begin{align}
\ket{\psi_0}=
\frac{1}{\sqrt{2}}\left(l_1^{1\dag} \psi_1^{2\dag} - l_1^{2\dag} \psi_1^{1^\dag}\right)
\frac{1}{\sqrt{2}}\left(l_2^{1\dag} \psi_2^{2\dag} - l_2^{2\dag} \psi_2^{1^\dag}\right)
\ket{0}
\end{align}
and has exactly one rishon per link. If the gauge-variant terms are suppressed by dissipation, then the single occupation of links is approximately conserved by the dynamics, which has the advantage of making rishon density--density interactions irrelevant (allowing to neglect the barred-out terms in Table~\ref{tab:scatteringMatrixElements}). 
The choice of one rishon per link corresponds to choosing a particular representation of U(2), although other possibilities (like, e.g., two rishons per link) are also allowed. In general, a one-rishon-per-link representation is possible for all U(N) QLMs (but not in the SU(N) case)~\cite{Banerjee2013sup,Wiese2013sup}. As can be seen from Fig.~3(b) of the main text, gauge invariance is violated for vanishing dissipation ($\kappa=0$), and restored for increasing $\kappa/J$, thus confirming the versatility of the dissipative protection scheme also for the non-Abelian U(2) quantum link model.

\subsection{Realization of noise constraints}

Now, we briefly describe how the noise sources required by Eq.~(1) of the main text can be added to the optical-lattice setup outlined in the previous section. In the following, we first describe the noise protection on individual building blocks and then discuss how the protection for the entire system can be constructed from these. 

Before proceeding to the specific example of the U(2) quantum link model, we stress that the Hermitian nature of the generators leads to the following additional freedom in the implementation: The Lindblad terms appearing in Eq.~(2) of the main text, $2G\rho G - G^2 \rho - \rho G^2\equiv \mD[G]\rho $ are invariant under a constant shift, i.e., $\mD[G+ {\rm const.} \times \mathbbm{1}]\rho=\mD[G]\rho$. This means that constants appearing in the generators can be ignored, since their presence is not important for the dissipative dynamics that we are interested in. In the following, we will exploit this fact several times to show how protection by classical noise can be achieved for the specific case of the U(2) model. To do so, we consider color-changing and color-preserving generators separately.

\subsubsection{Color-preserving generators $G_x$ and $G_x^{3}$}

The generators $G_x$ and $G_x^3$ are linear functions of fermion and rishon number operators and hence preserve the color degree of freedom. 
The corresponding terms in the noise Hamiltonian appearing in Eq.~(1) of the main text can be written as
\begin{align}
H_{\rm noise}^\prime 
&= \sqrt{2\kappa}\, \sum_x\xi_x^3(t)G_x^3 + \sqrt{2\kappa}\,\sum_x\xi_x(t)G_x^\prime\\
&= \sqrt{2\kappa}\, \sum_x\xi_x^3(t) 
\left[ N_x^1 - N_x^2 \right] 
+ \sqrt{2\kappa}\,\sum_x\xi_x(t) 
\left[ \psi_{x}^{\alpha\dagger}\psi_{x}^{\alpha}
-\frac{1}{2}(l_{x+1}^{\alpha\dagger}l_{x+1}^{\alpha}-r_{x}^{\alpha\dagger}r_{x}^{\alpha})
+\frac{1}{2}(l_{x}^{\alpha\dagger}l_{x}^{\alpha}-r_{x-1}^{\alpha\dagger}r_{x-1}^{\alpha})
\right]\,.
\end{align}
Here, $N_x^c=n_x^c+n_{l,x}^c+n_{r,x}^c$ is the total number of particles with color $c=1,2$ on building block $x$. In addition, $G_x^\prime$ denotes $G_x$ without the constant contribution, which we dropped due to its irrelevance as explained above.  Noise terms as in the above Hamiltonian can be realized by noisy AC-Stark shifts introduced by off-resonant lasers with intensity noise. Note that the generators involve atoms in several wells, which leads to a spatially correlated noise pattern. This can be achieved easily by using the same noise signal to control the AC-Stark shift on the involved wells, or even by using the same laser to address those wells. 

\subsubsection{Color-changing generators $G_x^{1}$ and $G_x^{2}$}

\begin{figure}
\centering
\includegraphics[width=0.9\textwidth]{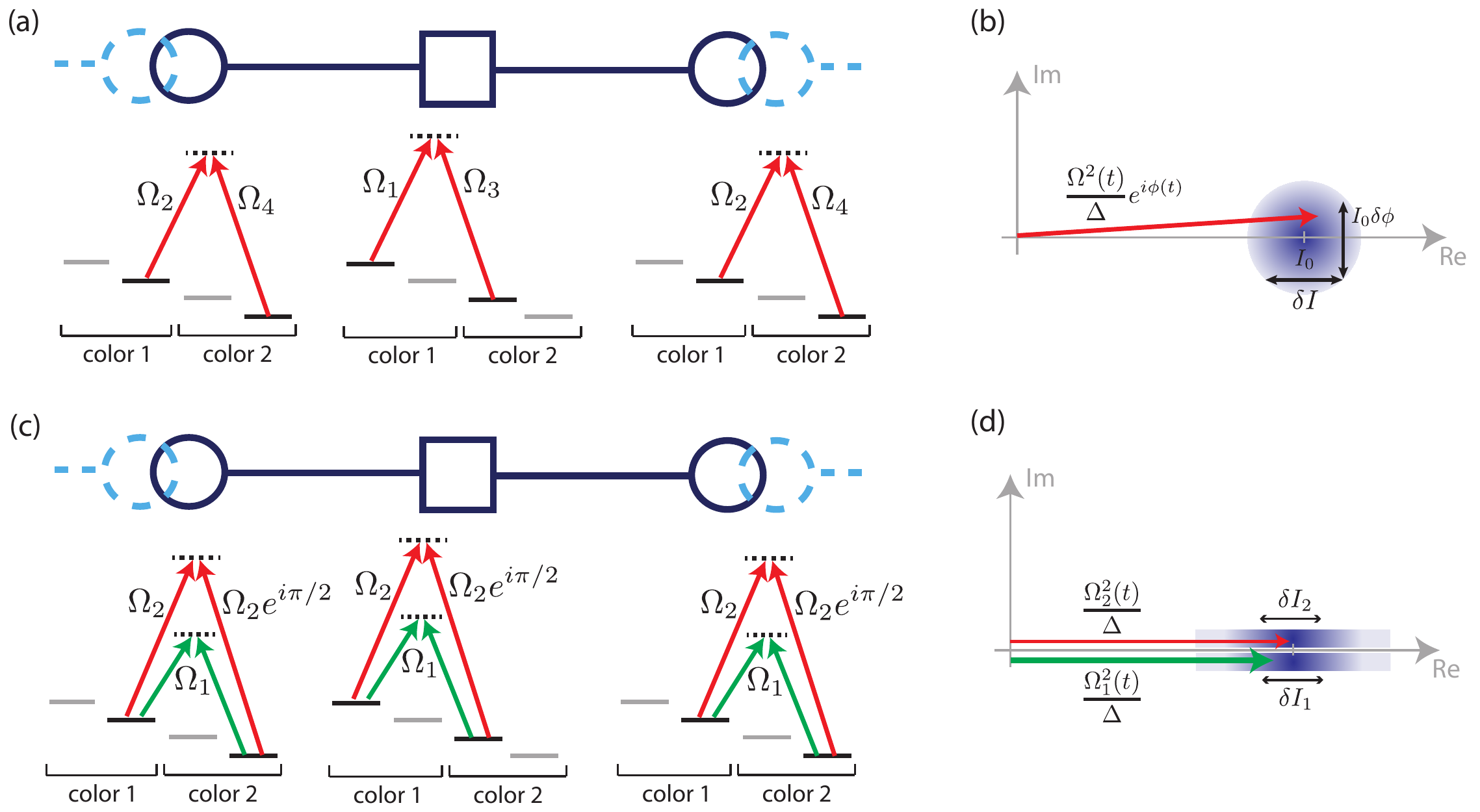}
\caption{Raman lasers for imposing the noise constraints associated with $G_x^{1,2}$. 
Upper panels: (a) Setup for generating $G_x^{1}$ and $G_x^{2}$ simultaneously using a single Raman transition. (b) The two required noise processes are realized by  uncorrelated fluctuations of intensity ($\delta I$) and phase ($\delta\phi$). 
Lower panels: (c) Setup employing one Raman transition per generator. (d) The two required noise sources are realized as independent intensity fluctuations $\delta I_i$, $i=1,2$, on the two transitions.
The grey levels are not trapped on the respective wells (boxes denote sites, and circles denote links, see~\figref{fig:implementation}). }
\label{fig:raman_lasers}
\end{figure}

To enforce the conservation laws corresponding to the remaining generators $G_x^1$ and $G_x^2$, we have to realize the noise Hamiltonian 
\begin{align}
H_{\rm noise}^{\prime\prime}
&=\sqrt{2\kappa}\,\sum_x\left[ \xi^1_x(t) G_x^1 + \xi_x^2(t) G_x^2 \right]\\
&=\sqrt{2\kappa}\,\sum_x \left[(\xi_x^1(t)+i\xi_x^2(t)) 
(\psi_x^{2\dag}\psi_x^{1} + l_{x}^{2\dag}l_{x}^{1} + r_{x}^{2\dag}r_{x}^{1})
+\hc\right]\,.
\label{eq:G12noise}
\end{align}
In the second line, we have combined the noise sources such that they appear as a single, complex valued noise signal $\xi_1+i\xi_2$. From the above expressions it is evident that the noise has to change the color degree of freedom. Since color is represented by internal levels in the setup proposed above, it is natural to employ Raman transitions for realizing such terms. A suitable arrangement of lasers is shown in \figref{fig:raman_lasers}(a), which is described by the following Hamiltonian:
\begin{align}
H_{\rm Raman}&=
 \frac{\Omega_1\Omega_3^*}{\Delta}\psi_x^{2\dag}\psi_x^{1}
+\frac{\Omega_2\Omega_4^*}{\Delta}l_{x}^{2\dag}l_{x}^{1}
+\frac{\Omega_2\Omega_4^*}{\Delta}r_{x}^{2\dag}r_{x}^{1}
+\hc\nonumber\\
&+\frac{\sabs{\Omega_1}^2}{\Delta} \psi_x^{1\dag}\psi_x^1
+\frac{\sabs{\Omega_3}^2}{\Delta} \psi_x^{2\dag}\psi_x^2
+\frac{\sabs{\Omega_2}^2}{\Delta} \left(r_{x}^{1\dag}r_{x}^{1}+l_{x}^{1\dag}l_{x}^{1}\right)
+\frac{\sabs{\Omega_4}^2}{\Delta} \left(r_{x}^{2\dag}r_{x}^{2}+l_{x}^{2\dag}l_{x}^{2}\right)\,.
\label{eq:G12Raman}
\end{align}
Here, the $\Omega_i$ are the complex Rabi frequencies on the corresponding transitions, and $\Delta$ is the detuning of the transitions from the excited states.
The first line in \eeqref{eq:G12Raman} yields exactly the desired terms displayed in Eq.~\eqref{eq:G12noise}, provided we have
$
 \frac{\Omega_1\Omega_3^*}{\Delta}
=\frac{\Omega_2\Omega_4^*}{\Delta}\propto\xi_x^1(t)+i\xi_x^2(t)
$.
This may be realized by choosing 
$\Omega_1=\Omega_2=\Omega(t)$, and
$\Omega_3=\Omega_4=\Omega(t) e^{i\delta\phi(t)}
$, where $\Omega(t)$ is taken to be real for simplicity, 
such that Eq.~\eqref{eq:G12Raman} becomes
\begin{align}
H_{\rm Raman}=
\frac{\Omega^2(t)}{\Delta}e^{i\delta\phi(t)}\left[\left(
 \psi_x^{2\dag}\psi_x^{1} + l_{xl}^{2\dag}l_{x}^{1} + r_{x}^{2\dag}r_{x}^{1}\right)
+\hc\right] \;\;+\frac{\Omega^2(t)}{\Delta} \left( N_x^1 + N_x^2 \right)\,.
\label{eq:raman}
\end{align}
For brevity, we introduce the ``intensity"  $I(t) \equiv \Omega^2(t)/\Delta\equiv I_0+\delta I(t)$, with mean $I_0$ and fluctuations of magnitude $\delta I(t)$. 
For small fluctuations $\delta I/I_0\ll1$ and  $\delta \phi\ll1$ as illustrated in \figref{fig:raman_lasers}(b), we can expand $I(t)e^{i\delta\phi}\approx I_0 + \delta I(t) + i I_0 \delta\phi(t) + \ldots $, such that \eeqref{eq:raman} becomes
\begin{align}
H_{\rm Raman}
\approx
I_0\left[
G_x^1 + N_x^1 + N_x^2 
\right]
+
\Big[
(\delta I(t) + i I_0 \delta\phi(t))
(\psi_x^{2\dag}\psi_x^{1} + l_{x}^{2\dag}l_{x}^{1} + r_{x}^{2\dag}r_{x}^{1})
+\hc
\Big]
+\delta I (t) \left[ N_x^1 + N_x^2 \right]\,.
\label{eq:HRaman2}
\end{align}
The first bracket only gives rise to trivial dynamics, as long as we are in the gauge-invariant subspace (where $G_x^1$ has eigenvalue 0) and the number of particles $N_x^1+N_x^2$ per building block is preserved, as it is the case for a system evolving according to $H^\prime_{\rm int}$  (see Table~\ref{tab:scatteringMatrixElements}). 
 The second bracket is the desired noise Hamiltonian displayed in \eeqref{eq:G12noise}, with $\xi_1+i\xi_2 \propto \delta I + iI_0 \delta\phi$, such that intensity and phase fluctuations coupling to $G_x^1$ and $G_x^2$, respectively. In addition, the third bracket is an undesired random stark shift, which is again irrelevant due to the conservation of $N_x^1+N_x^2$.

In order to realize the noise protection scheme outlined in the main text, the two signals $\delta\phi(t)$ and $\delta I (t)$ need to represent white noise sources. Since white noise is an idealization, we briefly describe how to choose the noise processes if we assume that they behave as Ornstein--Uhlenbeck processes \cite{Gardiner2000sup}
\begin{align}
d\delta I &= -i\gamma_I\, \delta I \, dt + \sqrt{D_I}\,dW_I\,, \\
d\delta \phi &= -i\gamma_\phi\, \delta \phi \, dt + \sqrt{D_\phi}\,dW_\phi\,,
\end{align}
with relaxation rates $\gamma_{I,\phi}$, diffusion constants $D_{I,\phi}$, and  uncorrelated Wiener increments $dW_{I,\phi}$ \cite{Gardiner2004sup}. The white-noise limit is reached when the correlation times $1/\gamma_{I,\phi}$ are much shorter than any other time-scale of the problem, in particular $1/\gamma_{I,\phi}\ll 1/\lambda$, with $\lambda$ the scale of the gauge-variant error terms. 
The diffusion constants can then be adjusted to meet the conditions for the small-noise expansion carried out above, i.e., $\lat\delta I^2 \rat =D_I/2\gamma_I\ll I_0^2$ and $\lat\delta \phi\rat=D_\phi/2\gamma_\phi\ll 1$, such that the second bracket in \eeqref{eq:HRaman2} realizes the desired noise Hamiltonian \eeqref{eq:G12noise}.

Averaging over the noise in \eeqref{eq:HRaman2} and taking the white-noise limit \cite{Gardiner2004sup} yields, in accordance with our earlier developments in Sec.\,\ref{sec:noise}, a master equation with two Lindblad terms:
\begin{align}
\label{eq:lindbladTerms}
\lat\dot\rho\rat \propto  \kappa_1 \mD[G_x^1 + N_x^1 + N_x^2]\lat\rho\rat + \kappa_2 \mD[G_x^2]\lat\rho\rat\,. 
\end{align}
where $\kappa_1=D_I/2\gamma_I^2$ and $\kappa_2=I_0^2D_\phi/2\gamma_\phi^2$. 
Since $N_x=N_x^1+N_x^2$ is a constant of motion under $H_{\rm int}^\prime$, we can finally replace $\mD[G_x^1 + N_x]\rightarrow\mD[G_x^1]$, as discussed above.
The coupling scheme presented here thus gives rise to the desired dissipative  coupling to the generators $G_x^{1,2}$. Together with the noisy AC-Stark shifts coupling to $G_x$ and $G_x^3$ outlined in the previous subsection one is thus able to realize the intended dissipative protection of gauge-invariant dynamics.

Finally, we point out an alternative for the realization of $H_{\rm noise}^{\prime\prime}$, which is depicted in \figref{fig:raman_lasers}(c,d): Instead of using a single Raman transition with amplitude and phase noise, one could realize \eeqref{eq:G12noise} by using a separate Raman transition for $G_x^{1}$ and $G_x^{2}$. This scheme works analogously to the one described above, but requires only intensity noise.

\subsubsection{Multiple building blocks}

If the gauge-variant perturbations contained in $H_1$ couple sites at arbitrarily long distances, noisy Raman beams and AC-Stark shifts should address all building blocks in an independent manner, with noise correlations only within a given building block. 
Such independent noise sources can be achieved by the high-resolution techniques developed in recent years, either by shining lasers beams onto the sample through holographic masks  \cite{Bakr2009sup,Weitenberg2011sup}, or by high-resolution microscopes such as employed in Refs.~\cite{Zimmermann2011sup,Bourgain2012sup}. 
In the latter case, microscopic dipole traps have been demonstrated that can be controlled independently and that have as little separation as less than $1\mu$m. 
Such focusing techniques should be sufficient to shine noise-modulated, focused Raman beams only on the few neighboring sites constituting a given building block. 

However, a drastic simplification will be possible if perturbations do not couple distant lattice wells. 
This will be the case in most practical implementations, since typical perturbations contained in $H_1$ act only locally, examples being nearest-neighbor collisions or tunnelings. 
Namely, if perturbations couple only neighboring building blocks $x$ and $x+1$, one can allow the noise on building block $x$ to be correlated to the one on $x+2$. 
Exploiting that neighboring building blocks are occupied by different species, one can then use two interleaved bundels of standing waves, one acting on species $\Phi$ and one on $\Psi$.  
The noisy AC-Stark shifts and Raman transitions generated by these global beams are then uncorrelated between building blocks $x$ and $x+1$, and thus suppress the nearest-neighbor perturbations $H_1$. 
In cases where longer-range perturbations become relevant, one may need to increase the periodicity after which the noise modulation is repeated, for example by applying the noise via global beams in a superlattice configuration. 
Therefore, since in most implementations perturbations will be of short range, one can circumvent the need for beams that act only on a few wells by global beam configurations. The numerical calculations in the next section demonstrate how, in the absence of long-range perturbations, such staggered noise sources can indeed replace the completely independent ones. 

In all cases, the correct coupling to the generators requires strictly speaking a step-like profile of the laser beams, with support only on a given building block. 
Realistically, and especially when using standing-wave configurations, one can expect deviations from this profile. Small deviations, however, prove to be irrelevant. 
Namely, a finite overlap of a beam to the next building block may be disregarded, since here a different species resides to which the laser will not couple. 
Further, an imperfect weight of the noise sources within a given building block will typically induce errors only at second order in the imperfection strength (as explained in the following section).

\subsection{Imperfections\label{sec:Imperfections}}

\begin{figure}[t]
\includegraphics[width=0.65\textwidth]{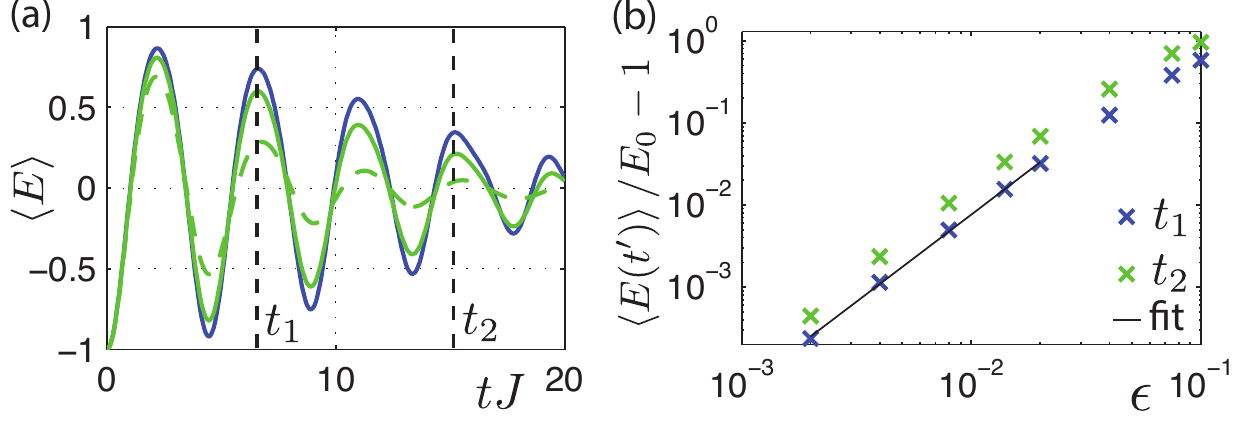}\caption{Influence
of imperfections on quench dynamics in a U(1) model (same scenario as in
Fig.\,2 of the main text). (a) Blue curve: perfectly engineered noise ($\kappa/J=10$, $\lambda/J=0.25$). Solid green curve: imperfect noise addressing with $\epsilon=0.05$ (see text). Dashed green curve: evolution under an onsite decay modeled by Lindblad terms $\gamma_0\sum_x(2\psi_x \rho \psi_x^\dag - \psi_x^\dag\psi_x \rho- \rho \psi_x^\dag\psi_x )$ with $\gamma_{0}/J=0.05$.  
(b) Scaling of the error due to imperfect noise addressing, estimated by the relative deviation of $\langle{E}(t^\prime=t_{1,2})\rangle$ from a perfect implementation of dissipative protection ($E_0$).
From the fitted straight line (black), we find that the error scales as $\sim \epsilon^{2.1}$.}%
\label{fig:imperfections}%
\end{figure}

Finally, we assess the influence of imperfections on the dissipative protection scheme. 
In an experimental realization of the scenario described in this work, the noise may not couple perfectly to the generators, 
but to slightly different operators $\tilde G_{x}^a = G_{x}^a + \epsilon f_{x}^a$, with $\epsilon\ll1$. 
Since the resulting dissipative terms in the master equation \eqref{eq:gn} read $\kappa(2\tilde G_x^a \rho \tilde G_x^a - \rho (\tilde G_x^a)^2 - (\tilde G_x^a)^2 \rho)$, the effects of such addressing errors scale, in general, linearly with $\epsilon$. 
However, the most common case will simply be a perturbed relative weight of the different terms in the generators, such as $f_{x}=\psi_{x}^{\dag}\psi_{x}$ for the U(1) model considered in the main text. 
In this case $[G_{x},f_{x}]=0$ and the effects of imperfect coupling scale as $\epsilon^{2}\kappa$, as long as the system remains in the gauge-invariant subspace. 
Requiring that $\epsilon$ be small compared to the scale $J$  imposes $\epsilon\ll\sqrt{J/\kappa}$ (instead of $\epsilon\ll J/\kappa$ as one would expect naively). 
In \figref{fig:imperfections}, we present numerical simulations of the master equation for the U(1) model including such imperfections. The blue curve in panel (a) displays the time evolution of the electric field for perfectly engineered noise, while the green curve represents the behavior for imperfect noise addressing. The quadratic scaling in $\epsilon$ is confirmed by the analysis presented in panel (b).

\section{Dissipative protection and scaling with the system size}

In view of possible applications in cold atoms systems, one of the key aspects of dissipative protection of Hilbert sub-space dynamics is the scaling properties as a function of the system size, or, in general, the number of degrees of freedom. In this section, we address this issue by means of exact numerics on small systems of different size. In particular, we look at the scaling of low-order correlations, such as order parameters. Let us briefly recall all fundamental ingredients of the dissipative protection scenario: 
\begin{itemize}
\item the total Hilbert space is separated into a gauge-invariant part $\mathcal{H}_\mathcal{P}$ and a gauge-variant part $\mathcal{H}_\mathcal{Q}$, with associated projection operators $\mathcal{P}$, $\mathcal{Q}$;
\item the Hamiltonian $H_0$ respects the separation between the two spaces, i.e., $H_0 \equiv \mathcal{P}H_0\mathcal{P}+\mathcal{Q}H_0\mathcal{Q}$;
\item the Hamiltonian $H_1$ mixes the two subspaces, i.e. $H_1 \equiv \mathcal{Q}H_1\mathcal{P}+\mathcal{P}H_1\mathcal{Q} + \mathcal{Q}H_1\mathcal{Q}$ (scale $\lambda$);
\item a set of classical noise sources $\xi_\alpha(t)$ protects $\mathcal{H}_\mathcal{P}$ by coupling to a set of Hermitian operators $G_\alpha$, defined by $G_\alpha |\psi\rangle=0$ if and only if $|\psi\rangle\in \mH_\mP$;
\item an initial state $|\psi_0\rangle\in \mH_\mP$;
\item an observable $M(t)$ that quantifies how the system leaves $\mathcal{H}_\mathcal{P}$ (this observable has close ties with $g^2(t)$ discussed in the main text in the case of gauge theories).
\end{itemize}

In the remainder of this section, we will show that the following two statements on the system dynamics hold:
{\it i)} On the one hand, the timescale associated with the exponential departure of the system dynamics from the protected Hilbert subspace as measured by the wave-function overlap $\mean{\mathcal{P}}(t)$ is expected to scale linearly with the system size, that is, the corresponding decay time would scale as $\tau_P\simeq \kappa/ ( N \lambda^2)$. The reason for this  is that  the number of finite matrix elements connecting the two subspaces $(\mH_\mP,\mH_\mQ)$ scales (in general) extensively with the system size.

{\it ii)} On the other hand, low-order correlations, and in particular order parameters, are not expected to follow this scaling. On the contrary, in the regime of protected dynamics, $t\ll \kappa/\lambda^2$, one expects that low-order correlation functions are not dramatically affected by $H_1$, since the latter only induces local changes on the system density matrix, while preserving global properties. The corresponding decay time for observables is then $\tau_{ob}=\kappa/\lambda^2$. This insensitivity to external perturbations (or imperfections) is just the reason why we care about "phases" in a condensed-matter phase diagram in the first place. 

\subsection{Model: Dissipative protection of a ferromagnet state}

\begin{figure}[t]
\centering
\includegraphics[width=0.6\columnwidth]{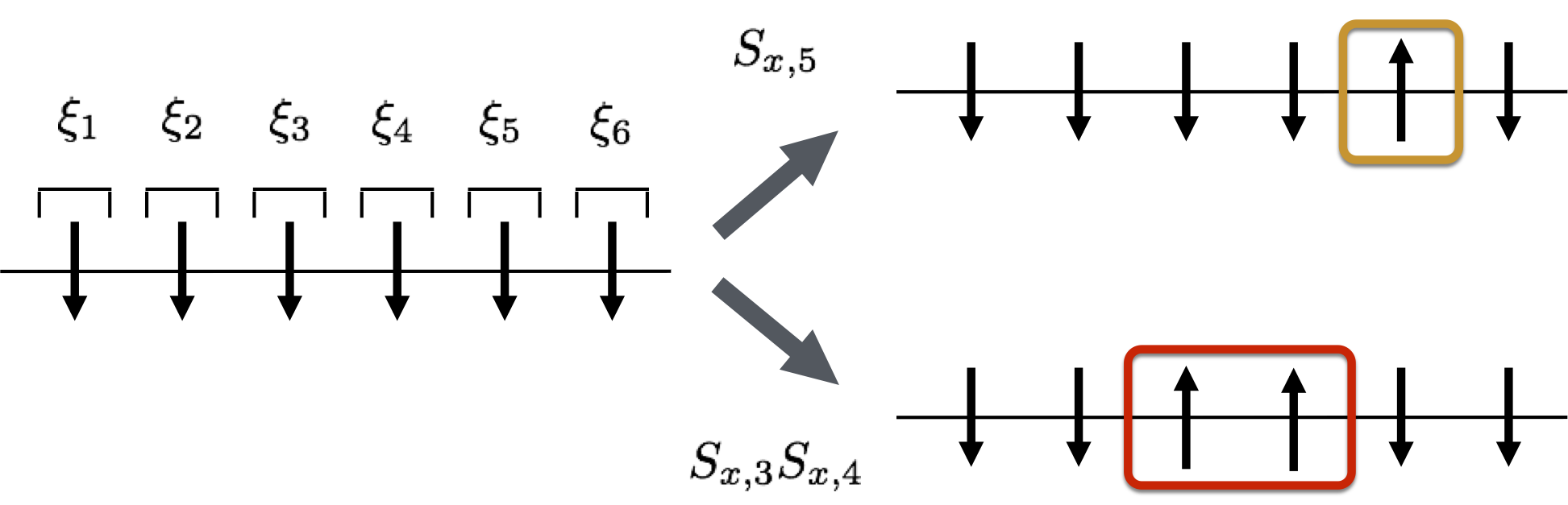}
\caption{Cartoon of the spin model's microscopic dynamics. Left side: a ferromagnetic state $|F\rangle$ is dissipatively protected by a set of noise sources $\xi_j(t)$ which pin the local magnetic field. Right panel: the system dynamics induced by $H_1$ drives the system out of the protected subspace $\mathcal{H}_\mathcal{P}$ by flipping spins on either single sites (due to $\Delta\neq0$, upper cartoon) or on nearest-neighbor (due to interactions in the $x$-spin component, lower cartoon). }
\label{fig:cart}
\end{figure}

The above qualitative observations require in general quantitative techniques to be benchmarked. However, performing unbiased numerical simulations of the LGTs presented in the main text is very challenging because of constraints in the system sizes achievable due to both the structure of the gauge theory, and the complexity of a full master equation treatment. 
Therefore,  we consider  a simplified scenario in which the scaling of the protection mechanism is expected to behave similarly as for the LGTs we are interested in. In particular, we study the dynamics of a circular chain consisting of  $N$ spin-1/2 particles under dissipative protection of a ferromagnetic state. This way, we have $\dim\mH_\mP=1$ and $\dim\mH_\mQ\simeq 2^N$, such that the ratio of subspace dimensions is far less favourable than for the LGT models presented in the main text.
To be more specific, the ingredients for the model are as follows:
\begin{itemize}
\item as the protected space, we make the extreme choice of choosing a single state, namely the $\downarrow$-ferromagnet: 
\begin{equation}
|F\rangle = \prod_{j=1}^{N}\otimes|\downarrow\rangle_j, \quad \mH_\mP = \{ |F\rangle \}.
\end{equation}
This state will also serve as the initial condition for the quantum dynamics, and defines a relevant order parameter to be preserved, $M(t) = \sum_{j} \langle s^z_j\rangle_t/N$, where $M(0) = -0.5$;
\item all other states in the spin chain define the subspace $\mathcal{H_Q}$;
\item as the Hamiltonian $H_0$ that preserves the initial state we choose a general Ising-interaction:
\begin{equation}
H_0 = h\sum_{i}s^z_i + J_z\sum_{<i,j>} s^z_is^z_{j}\,;
\end{equation}
\item $H_1$ can have many possible forms; we choose
\begin{equation}
H_1 = \Delta\sum_{i}s^x_i + J\sum_{<i,j>} s^x_is^x_{j}\, ,
\end{equation}
such that the coherent part $H_0+H_1$ of the model cannot be reduced to single-particle dynamics;
\item regarding the dissipative protection of the initial state, we choose two different noise configurations, which are in close analogy to the LGTs presented in the main text:
\begin{enumerate}
\item a pair of noise sources keeping the magnetisation on the even/odd sub lattices fixed:
\begin{equation}
H_{\textrm{noise}} = 
   \sqrt{\kappa}\,\xi_{\textrm{even}}(t)\sum_{i\, \textrm{even}}s^z_i 
+  \sqrt{\kappa}\,\xi_{\textrm{odd}}(t)\sum_{i\, \textrm{odd}}s^z_i\,,
\end{equation}
which  corresponds to two jump operators $C_\textrm{even/odd} = \sum_{i\, \textrm{even/odd}}s^z_i$
\item a set of $N$ noise sources:
\begin{equation}
H_{\textrm{noise}} = \sqrt{\kappa}\,\sum_{i}\xi_{i}(t) s^z_i\,, 
\end{equation}
which corresponds to $N$ jump operators $C_{i} =s^z_i$
\end{enumerate}
\end{itemize}
The various effects of the dissipative and Hamiltonian dynamics are illustrated qualitatively in Fig.~\ref{fig:cart}: While the noise sources pin the local magnetization in the down state (left panel), the Hamiltonian terms in $H_1$ try to flip spins and thereby destroy the ferromagnetic order (right panel).

In Figs.~\ref{fig:fid}, ~\ref{fig:magn} and \ref{fig:magn2} we present results obtained for up to $N=10$ spins by numerical solution of the master equation as formulated in Eqs.\,\ref{eq:MEnoise}-\ref{eq:gn}, with the role of the $G_x^a$ played by the $C_i$ or $C_{\textrm{even/odd}}$.  We have investigated various parameter regimes in terms of both noise and energy scales, however, for the sake of concision, we only present results for  $h=0.5, \Delta=1.5$ and  $J_z=J =1.0$ ($J$ also provides the energy scale reference in the figures). Note that Figs.~\ref{fig:fid} and ~\ref{fig:magn} show results for  $N$  noise sources, while Fig.~\ref{fig:magn2}  concerns both noise configurations (a complete overview will be presented elsewhere \cite{dalmonte_un_sup}). Notice that it is important to have all energy scales in $H_1$ comparable or larger than the ones in $H_0$ in order to rule out a possible energetic protection.

\subsection{Numerical Results}

\subsubsection{Population of the protected subspace}

\begin{figure}[t]
\centering
\includegraphics[width=0.96\columnwidth]{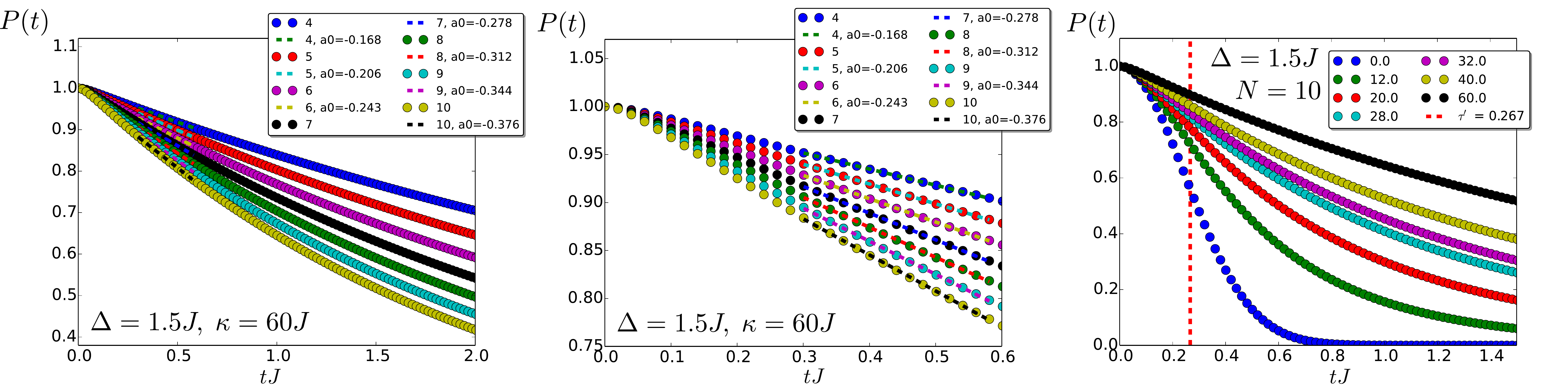}
\caption{Population in the protected subspace after an interaction quench, $N$ sources. Left panel: fixed $\kappa$, various $N$ (indices in the legend). Middle panel: zoom-in on the left panel for short timescales. Right panel: fixed $N$, various $\kappa$ (indices in the legend). The line at $\tau'\propto \lambda^2/\kappa$  represents an indicative timescale for fidelity decay at $\kappa=60$. In all panels, points represent the numerical results; In the first two panels, dashed lines indicate linear fits: the linear coefficient obtained out of the fit is indicated in the legend as $a_0$. }
\label{fig:fid}
\end{figure}

As a starting point, we benchmark our prediction {\it i)} about the population in the protected space, which for the present model is nothing but the overlap with the only state in $\mH_\mP$, that is:
\begin{equation}
P(t) = \langle \mathcal{P} \rangle(t), \quad \mathcal{P}= |F\rangle\langle F|.
\end{equation}
Fig.~\ref{fig:fid} presents typical results for the scaling of this figure-of-merit:
In the right panel, we plot the scaling of $P(t)$ as a function of time for different protection strengths $\kappa$. In the first and second panel, we present typical results in the protected regime, $\kappa = 60 \gg 1.5 =\lambda= \max[\Delta, J]$, 
as a function of the system size $N$. $P(t)$ deteriorates with the system size, and moreover, as evidenced in the second panel, the short-time dynamics is characterised by a linear decay with timescale $\tau\propto 1/ N$. 
All results confirm our expectations from {\it i)}, indicating that for this simple model, extracting a proper large-$N$ limit from limited size samples is feasible.

\subsubsection{Many-body observables}

\begin{figure}[h]
\centering
\includegraphics[width=0.96\columnwidth]{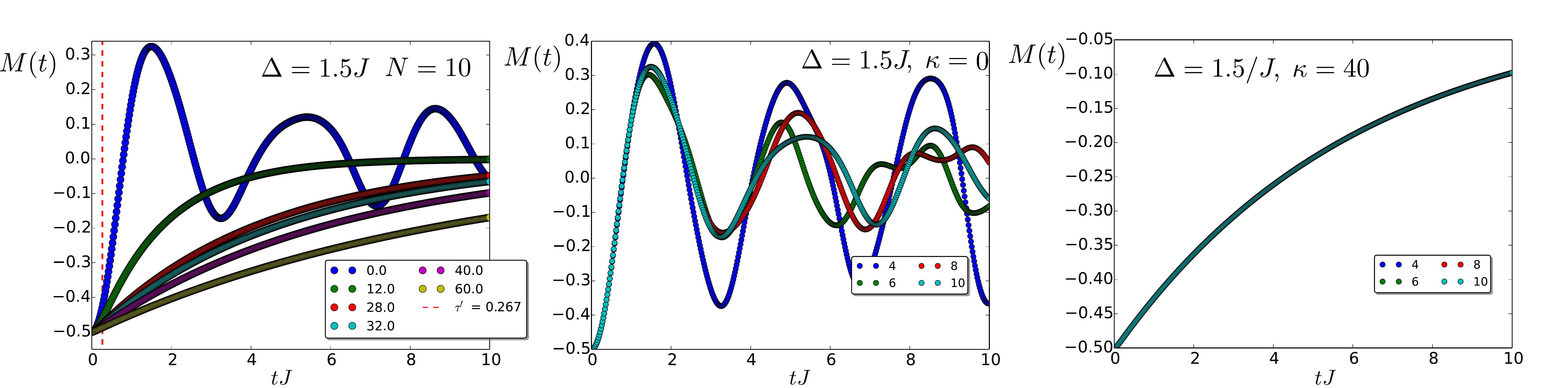}
\caption{Magnetization scaling after an interaction quench, $N$ sources. Left panel: fixed $N=10$, various $\kappa$. Middle and right panels: magnetisation scaling in the unprotected (middle) and protected regime (right) for different system sizes. In the first panel, the dashed line at $\tau'\propto \lambda^2/\kappa$ represent an indicative timescale for fidelity decay for $\kappa=60$.
}
\label{fig:magn}
\end{figure}

We now turn to point {\it ii)} to address the behaviour of low-order correlation functions during the dissipatively protected dynamics. An overview of the behavior of the ferromagnetic order parameter $M(t)$ is presented in Fig.~\ref{fig:magn}. The first panel shows the scaling of $M(t)$ as a function of time for different protection strengths: as expected, the behaviour is very similar to the one reported in the main text for both Abelian and non-Abelian gauge theories. For comparison, the second panel describes the unprotected case: there, magnetic order is lost at all system sizes on a short timescale. The third panel shows the magnetisation scaling in the protected case, $\kappa= 40$. Here, all curves for different system sizes practically overlap -- appreciable differences appearing only far out of the protected regime, defined by $t\ll \kappa/\lambda^2$. 

\begin{figure}[h]
\centering
\includegraphics[width=0.96\columnwidth]{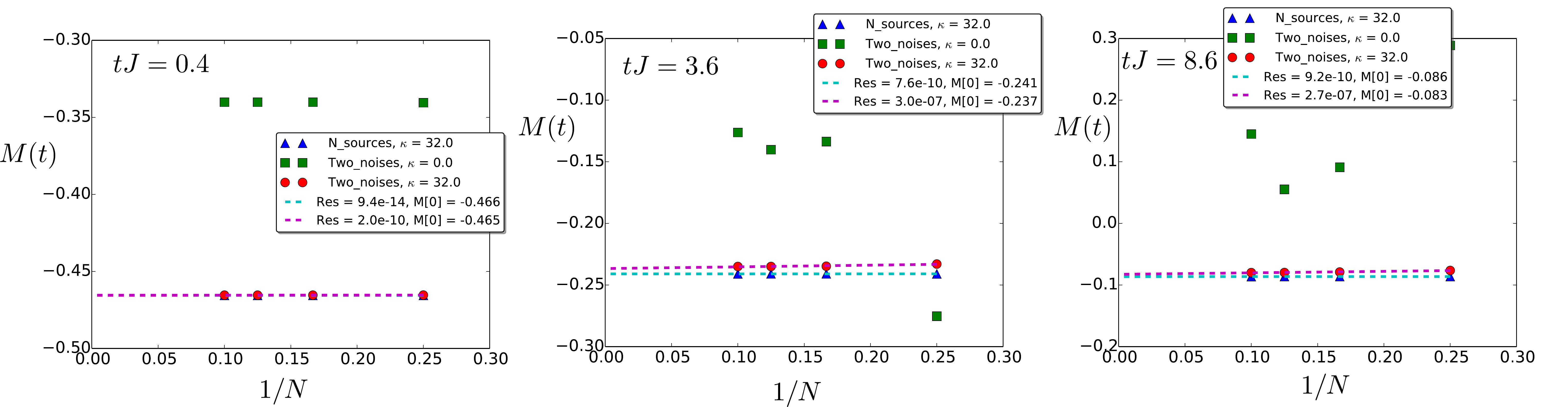}
\caption{Magnetization scaling after an interaction quench as a function of the inverse system size, for either unprotected case (green squares), N-noise protection (blue triangles) and 2-noise protection (red circles). From left to right: snapshots at different times $tJ=0.4, 3.6, 8.6$. Dashed lines are linear fits as a function of $1/N$: in the corresponding legend, the residue and the extrapolated thermodynamic value are indicated (Res and $M[0]$, respectively). In all cases, the protected dynamics leads to an order parameter which is basically independent on the system size, and shows very weak dependence even on long-time scales (right panel). The unprotected dynamics strongly depends on the system size. }
\label{fig:magn2}
\end{figure}

A more quantitative analysis of the finite-size scaling of $M$ for different timescales and protections is provided in Fig.~\ref{fig:magn2}. There, we plot the scaling of $M(t)$ as a function of $1/N$ in order to check the efficiency of the protection scheme in the thermodynamic limit. For very short timescales $tJ =0.4$, the magnetisation is almost independent of the system size, and its scaling is very well described by a linear scaling with finite and large magnetisation in the thermodynamic limit. The unprotected case seems also to be almost system size independent.

Moving to longer timescales, the protected cases still scale as expected, namely, there is no appreciable dependence on the system size: the extrapolated value of the order parameter in the thermodynamic limit differs from the finite-size one only at the percent level or lower. On the other hand, the unprotected case shows strong system-size oscillations, as expected from the picture given in Fig.~\ref{fig:magn}. Another interesting feature is that the two-source scheme (red circles) performs almost as good as the $N$-source case (blue triangles), both in terms of absolute values than in terms of scaling properties. 

In summary, all results point toward an order-parameter decay-time independent of the system size, as argued in {\it ii)}, for both kind of protection schemes. This indicates that the protection scheme devised does not suffer from $N$-scaling properties, even in the worst case scenario of a very small protected subspace investigated here. 
As in the LGT case the Hilbert space has always a better scaling compared to the spin model discussed here, we expect system-size independence to be very robust for the LGTs as well.


\begin{thebibliography}{53}
\bibitem{footnoteg0}
  \BibitemOpen
  \bibinfo {note} {For the purpose of illustrating the basic principle, we set
  the field coupling $\tilde{g}=0$.}\BibitemShut
  {Stop}%
\end{thebibliography}
\end{document}